\documentclass[final,5p,times,twocolumn]{elsarticle}

\usepackage{graphicx,color}
\usepackage{amsmath,amssymb,bm}

\def\Xint#1{\mathchoice
   {\XXint\displaystyle\textstyle{#1}}%
   {\XXint\textstyle\scriptstyle{#1}}%
   {\XXint\scriptstyle\scriptscriptstyle{#1}}%
   {\XXint\scriptscriptstyle\scriptscriptstyle{#1}}%
   \!\int}
\def\XXint#1#2#3{{\setbox0=\hbox{$#1{#2#3}{\int}$}
     \vcenter{\hbox{$#2#3$}}\kern-.5\wd0}}

\def\dashint{\Xint-}

\journal{Physics Letters B}

\begin{document}

\begin{frontmatter}
\title{Implicit vs Explicit Renormalization and Effective Interactions}

\author{E. Ruiz Arriola} 
\ead{earriola@ugr.es} 
\address{Departamento
  de F\'isica At\'omica, Molecular y Nuclear and Instituto Carlos I de
  Fisica Te\'orica y Computacional, \\ Universidad de Granada, E-18071
  Granada, Spain}

\author{S. Szpigel} 
\ead{szpigel@mackenzie.br}
\address{Faculdade de Computa\c c\~ao e Inform\'atica, 
Universidade Presbiteriana Mackenzie, Brazil}

\author{V. S. Tim\'oteo} 
\ead{varese@ft.unicamp.br}
\address{Grupo de \'Optica e Modelagem Num\'erica (GOMNI), Faculdade de Tecnologia, Universidade Estadual de Campinas - UNICAMP, Brazil}

\begin{abstract}
Effective interactions can be obtained from a renormalization group
analysis in two complementary ways. One can either explicitly
integrate out higher energy modes or impose given conditions at low
energies for a cut-off theory. While the first method is numerically
involved, the second one can be solved almost analytically.  In both
cases we compare the outcoming effective interactions for the two
nucleon system as functions of the cut-off scale and find a strikingly
wide energy region where both approaches overlap, corresponding to
relevant scales in light nuclei $\Lambda\lesssim 200 {\rm MeV}$. This
amounts to a great simplification in the determination of the effective
interaction parameters.
\end{abstract}

\end{frontmatter}

\section{Introduction}

Since half a century ago the idea of effective interactions has been
strongly pursued after the pioneering works by
Goldstone~\cite{Goldstone:1957zz}, Moshinsky~\cite{Moshinsky195819}
and Skyrme~\cite{Skyrme:1959zz}. They suggested to use this notion to
cut down the complexity of the Nuclear Many Body Problem due to strong
short range repulsion which arises when nucleon-nucleon (NN)
interactions are probed at sufficiently high energies. Effective
interactions were profitably exploited in the mid
70's~\cite{Vautherin:1971aw} and have reached a high degree of
sophistication (for a review see e.g.~\cite{Bender:2003jk}). A very
recent compilation of parameters is given in Ref.~\cite{Dutra:2012mb}
displaying a huge diversity, somewhat reflecting the disparate
phenomena which are used to fix the effective Hamiltonian, but
remarkably exhibiting no link to the fundamental two-body
interaction. In a recent work~\cite{Arriola:2010hj} (see
also~\cite{Nakamura:2004ek}) a model independent and implicit way of
determining the effective interactions from NN low energy scattering
data has been suggested.  They depend on the minimal de Broglie
wavelength between nucleons in a finite nucleus, a trend consistent
with fitting coarse grained NN interactions~\cite{NavarroPerez:2011fm}
to fixed upper center of mass (CM) momenta~\cite{NavarroPerez:2012qr}.

In the last decade there has been an intense reformulation of the
nuclear many body problem inspired by the Wilsonian renormalization
group ideas providing an alternative approach to the determination of
effective
interactions~\cite{Bogner:2001gq,Bogner:2003wn,Bogner:2006pc} (for
reviews see e.g.~\cite{Coraggio:2008in,Bogner:2009bt,
  Furnstahl:2013oba} and references therein) and their
characterization as finite cut-off
counterterms~\cite{Holt:2003rj}. This framework takes advantage of the
proper momentum scale resolution or cut-off $\Lambda$, separating
explicitly what degrees of freedom and interactions behave dynamically
below that scale. The requirement that observables should be cut-off
independent determines the implicit $\Lambda$ dependence of the
effective interaction. A direct and explicit way to achieve such
interaction uses the Similarity Renormalization Group (SRG) method
with a block-diagonal generator whence an effective hermitean
phase-equivalent interaction is derived~\cite{Anderson:2008mu}.

In the present paper we analyse the Block-Diagonal Similarity
Renormalization Group (BD-SRG) scheme~\cite{Anderson:2008mu} as
applied to the two body problem. This allows to implement by a
continuous and unitary evolution in a momentum-dimension auxiliary
parameter $\lambda$, referred to as the SRG-cutoff, a block-diagonal
separation of the Hilbert space in two orthogonal (decoupled)
subspaces ${\cal H}= {\cal H}_P \oplus {\cal H}_Q$ which are below or
above $\Lambda$ respectively. The evolution runs from $\lambda=\infty$
(the ultraviolet limit) to $\lambda=0$ (the infrared limit) and
interpolates between a {\it bare} Hamiltonian, $H_{\lambda=\infty}$,
and the block-diagonal one $H_{\lambda=0}$ in a unitary way
$H_{\lambda=0}=U H_{\lambda=\infty} U^\dagger$. This is the unitary
implementation~\cite{Anderson:2008mu} to all energies of the
previously proposed $V_{\rm low~k}$-approach~\cite{Bogner:2001gq}
where the higher energy states are missing, and in practice a free
theory was assumed above the energy determined by the momentum cut-off
$\Lambda$, hence generating a truncation error. For the rest of the
paper we will refer to this $\Lambda$ as the $V_{\rm low~k}$-cutoff to
be identified with the block-diagonal SRG one. We emphasize that a
complete Hilbert space separation corresponds to the limit $\lambda
\to 0$.

Although this block-diagonal scheme solves the problem as a matter of
principle, SRG equations are differential equations in the SRG-cut-off
$\lambda$ for unbound operators defined on the Hilbert space, and they
have only been solved exactly for simple
cases~\cite{Szpigel:1999gf}. For most cases however, SRG equations
must be numerically posed on a finite $N-$dimensional momentum grid,
$p_n$, and the differential equations require a further grid in the
SRG-cut-off $\lambda_i$ which introduces two infrared resolution
scales $\Delta p_n$ and $\Delta \lambda_i $. In the BD-SRG equations
$\Lambda$ takes values on the momentum grid $p_n$. The interplay among
these scales makes the limit $\lambda \le \Delta p, \Lambda$
numerically stiff and computationally expensive.  We will show that
this infrared behaviour is best reproduced by directly using low
energy scattering data in the continuum and, most remarkably, that
effective interactions are accurately determined this way in a wide
cut-off range.

\section{Bare and effective interaction}

We review briefly the renormalization problem for the two-nucleon
system from a Wilsonian point of view to introduce our notation in a
way that our results can be easily stated ( see
e.g. Ref.~\cite{Harada:2005tw} for an alternative set up). To motivate
the discussion let us consider NN scattering, where one solves the
Lippmann-Schwinger (LS) equation for the {\it bare} potential $V$.
Taking the case of S-waves we have for the half-off-shell $K-$matrix,
\begin{eqnarray}
K(p',p) = V(p',p) + \frac{2}{\pi} \dashint_0^\infty dq 
\frac{q^2 V(p',q)}{p^2-q^2} K(q,p)
\label{eq:LS-mom-bare} 
\end{eqnarray}
where $K(p',p)$ is the reaction matrix which relation to the
phase-shifts is given by
\begin{eqnarray}
\frac{\tan \delta(p)}{p}=-K(p,p)  
\label{eq:K-mom-bare} 
\end{eqnarray}
The
effective interation $V_{\Lambda}(p',p)$ corresponds to a
self-adjoint operator, $V_\Lambda(p',p)  = V_\Lambda(p,p')^*$, 
acting in a reduced model Hilbert space with $p,p'
\le \Lambda$ and fulfills 
\begin{eqnarray}
K_\Lambda (p',p)  = V_\Lambda (p',p) + \frac{2}{\pi} \dashint_0^\Lambda dq \frac{q^2V_\Lambda (p',q)}{p^2-q^2} 
K_\Lambda (q,p) \, .
\label{eq:LS-mom-eff} 
\end{eqnarray}
Using the similar definition of Eq.~(\ref{eq:K-mom-bare}) we get  
\begin{eqnarray}
\delta_{\Lambda} (p) = \delta (p) \Theta(\Lambda-p) \, . 
\label{eq:del-lowk}
\end{eqnarray}
The idea is that by using this truncation one can work in a smaller
space, without explicit reference to high energy states. This does not
provide a unique definition of the effective interaction, so an
auxiliary condition must be specified. In the original $V_{\rm low k}$
approach~\cite{Entem:2007jg} the half-off shell T-matrix was fixed to
the bare one, a procedure which did not guarantee a self-adjoint
operator, and hence a subsequent hermitization procedure was
required. In the BD-SRG approach~\cite{Anderson:2008mu} the
hermiticity is preserved along the SRG evolution.

The SRG method does not specify what the bare interaction should be
and is usually taken as a realistic potential which fits NN data up to
pion-production threshold, $\Lambda \lesssim \sqrt{m_\pi M_N} \sim 400
{\rm MeV}$ . This introduces a long high momentum tail due to the
short range repulsion which complicates the numerical convergence when
solving the SRG flow equations. For illustration purposes we take as
the {\it bare} interaction the simple separable potential for the NN
S-waves
\begin{eqnarray}
V_\alpha (p, p') = C_\alpha g_\alpha (p') g_\alpha(p) \qquad \alpha= ^1S_0, ^3S_1 
\label{gaupot}
\end{eqnarray}
leading to the phase-shifts
\begin{eqnarray}
 p \cot \delta_\alpha (p) &=& - \frac{1}{V_\alpha (p,p)} \left[1- \frac{2}{\pi}
   \dashint_0^\infty dq \frac{q^2}{p^2-q^2} V_\alpha(q,q) \right] \nonumber \\ 
&=& -\frac1{\alpha_0} +
 \frac12 r_0 p^2 + v_2 p^4 + \dots
\label{eq:ERE}
\end{eqnarray}
where in the last line a low momentum Effective Range Expansion  
(ERE) has been carried out and the scattering length $\alpha_0$, the
effective range $r_0$ and the $v_2$ parameter have been introduced.
Parameters in Eq.~(\ref{gaupot}) are adjusted to reproduce $\alpha_0$
and $r_0$ which for a gaussian form factor $g_\alpha (p) = e^{-p^2 /L^2}$ 
are listed in Table~\ref{tab:toy}. 
\begin{figure}[tbc]
\begin{center}
\includegraphics[height=5cm,width=7cm]{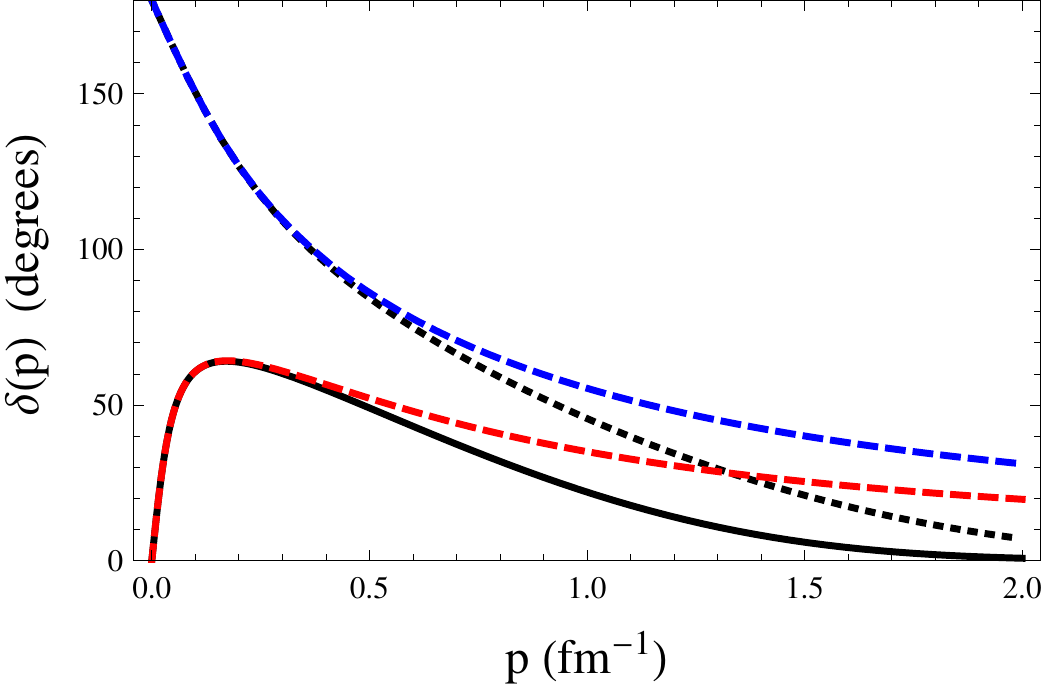}
\end{center}
\caption{$^1S_0$ (solid) and $^3S_1$ (dotted) phase-shifts in degrees for the
  separable potential and compared with the effective range expansion
  to second order (dashed) as a function of the CM momentum (in fm$^{-1}$).
}
\label{fig:phase-LS}
\end{figure}
The resulting phase-shifts are presented in Fig.~\ref{fig:phase-LS}
together with the ERE results, which reproduce well data up to $p \le
\Lambda_{\rm ERE} \sim 100 {\rm MeV}$. While they only resemble NN phase-shifts
of the most recent Partial Wave Analysis~\cite{Perez:2013mwa} at low
momenta, these two channels illustrate Levinson's theorem that
$\delta(0)-\delta(\infty)=n \pi $ with $n$ the number of bound states
and $n_{^1S_0}=0$ and $n_{^3S_1}=1$.  The pole of the $^3S_1$
scattering amplitude at $p=i \gamma= i 0.2314 \, {\rm fm}^{-1}$ gives
a satisfactory deuteron binding energy $E_d=-\gamma^2/M = -2.22 {\rm
  MeV}$.

\begin{table}[htb]
\begin{tabular}{|c|c|c|c|c|}
\hline 
Parameter  & $\alpha_0 $  &  $r_0 $ & $ C $  & $L $ \\
Units & $({\rm fm})$ &  $({\rm fm})$ &  $({\rm fm})$ &  $({\rm fm}^{-1})$ \\    
\hline 
$^1S_0$   &
-23.74  & 2.77 & -1.9158  &  0.6913 \\ 
$^3S_1$   &
5.42  & 1.75 &-2.3006  &  0.4151 \\ 
\hline
\end{tabular}
\caption{Model parameters for the gaussian separable potential $V_\alpha (p',p)=C_\alpha e^{-(p^2+p'^2)/L_\alpha^2}$ used in the calculations.}
\label{tab:toy}
\end{table}

\section{Explicit Renormalization: Block diagonal evolution}

The SRG method developed by Glazek and
Wilson~\cite{Glazek:1993rc,Glazek:1994qc} and independently by
Wegner~\cite{Wegner200177} (for a review see
e.g.~\cite{Kehrein:2006ti}) is based on a non-perturbative flow
equation that governs the unitary evolution of a hamiltonian $H=T_{\rm
  rel}+V$ with a flow parameter $s$ that ranges from $0$ to $\infty$,
\begin{equation}
\frac{d H_s}{ds}=[\eta_s,H_s]\; ,
\end{equation}
\noindent
where $\eta_s=[G_s,H_s]$ is an anti-hermitian operator that generates
the unitary transformations. We take the Block-diagonal SRG
generator~\cite{Anderson:2008mu} given by
\begin{eqnarray}
 G_{s}=H_{s}^{\rm BD}
  \equiv \begin{pmatrix}
    PH_{s}P  & 0 \\ \\
    0  & QH_{s}Q
  \end{pmatrix} \;.
\end{eqnarray}
\noindent
where $P$ and $Q=1-P$ are projection operators. The flow parameter $s$
has dimensions of $[{\rm energy}]^{-2}$ and in terms of a similarity
cutoff $\lambda$ with dimension of momentum is given by the relation
$s=\lambda^{-4}$.  The flow equation is to be solved with the boundary
condition $H_s |_{_{s \rightarrow s_0}} \equiv H_{s_0}$. Using that
$T_{\rm rel}$ is independent of $s$, we obtain
\begin{equation}
\frac{d V_s}{ds}=[\eta_s,H_s]\; . 
\label{eq:SRG}
\end{equation}
In a partial-wave
relative momentum space basis, the projection operators are determined
in terms of a momentum cutoff scale $\Lambda$ that divides the
momentum space into a low-momentum $P$-space ($p < \Lambda$) and a
high-momentum $Q$-space ($p > \Lambda$), 
\begin{eqnarray}
P \equiv \theta(\Lambda - p) ; \; Q \equiv \theta(p -\Lambda) \; .
\label{eq:PQ-sharp}
\end{eqnarray}
The potential $V_s$ can be written as,
\begin{eqnarray}
 V_{s}
  \equiv \begin{pmatrix}
    PV_{s}P  & PV_{s}Q \\ \\
    QV_{s}P  & QV_{s}Q
  \end{pmatrix} \;.
\end{eqnarray}
By choosing the block-diagonal generator, the matrix-elements inside
the off-diagonal blocks $PV_{s}Q$ and $QV_{s}P$ are suppressed as the
flow parameter $s$ increases (or as the similarity cutoff $\lambda$ decreases), such that the hamiltonian is driven to a block-diagonal
form,  
\begin{eqnarray}
\lim_{\lambda \to 0} V_\lambda = P V_{\rm low k} P + Q V_{\rm high k} Q =     
\begin{pmatrix}
V_{\rm low \, k}      & 0 \\ \\
    0  & V_{\rm high\, k}
  \end{pmatrix} 
\end{eqnarray}
Thus, in the limit $\lambda \rightarrow 0$ the $P$-space and the
$Q$-space become completely decoupled. Thus, while unitarity implies
$\delta_\lambda(p)= \delta(p)$ for any $\lambda$ one has
\begin{eqnarray}
\lim_{\lambda \to 0} \delta_\lambda (p) = 
\delta_{\rm low k} (p) 
+ \delta_{\rm high k} (p) 
\end{eqnarray}
where $\delta_{\rm low k} (p)=\delta(p) \theta(\Lambda-p) $ and 
$ \delta_{\rm high k} (p)= \delta(p) \theta(p-\Lambda)$ are the
phase shifts of the $V_{\rm low \, k} $ and $ V_{\rm high\, k} $
potentials respectively (see  Eq.~(\ref{eq:del-lowk})). 
 
\begin{figure*}[ht]
\begin{center}
\includegraphics[width=5.5cm]{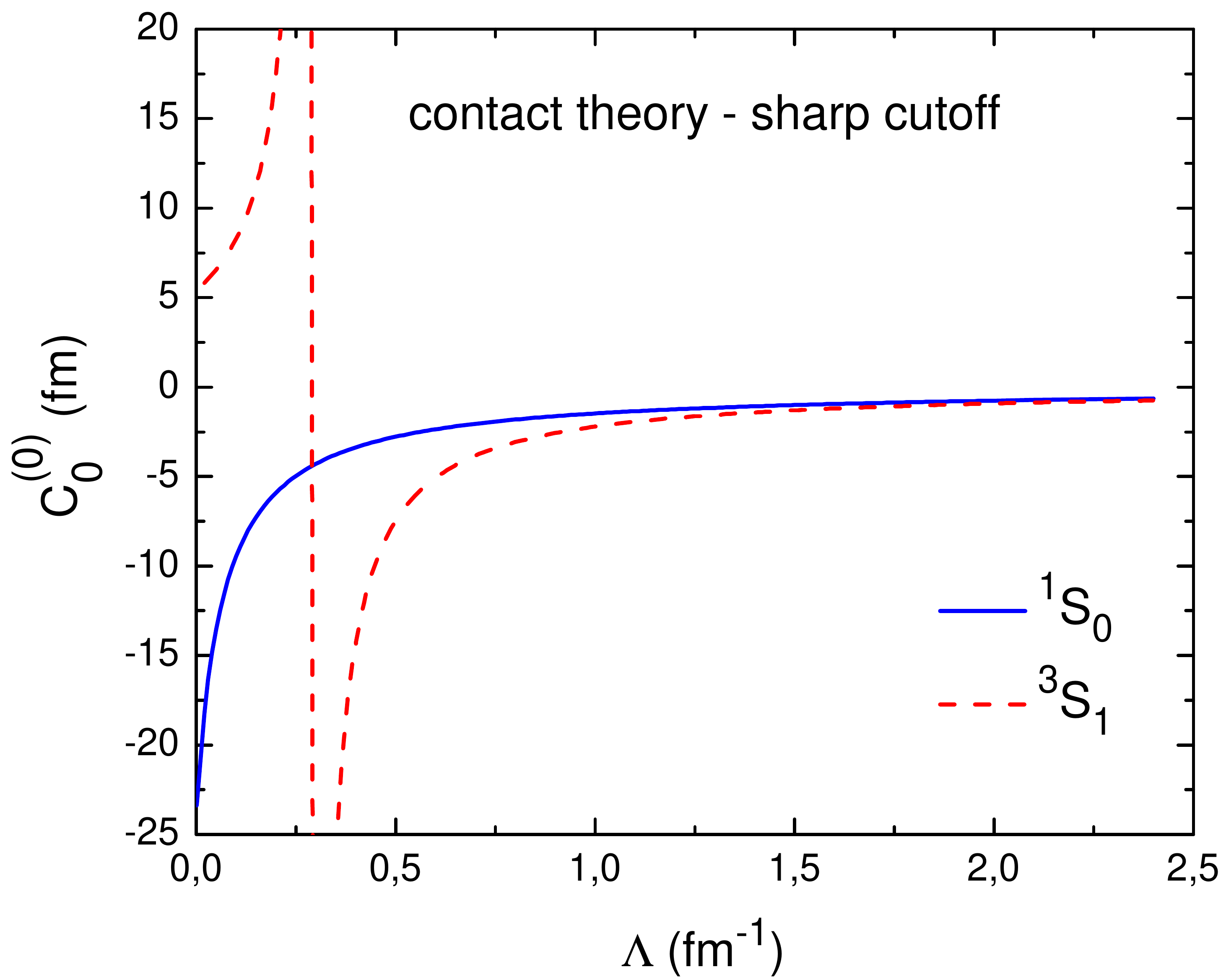}\hspace{0.2cm}
\includegraphics[width=5.5cm]{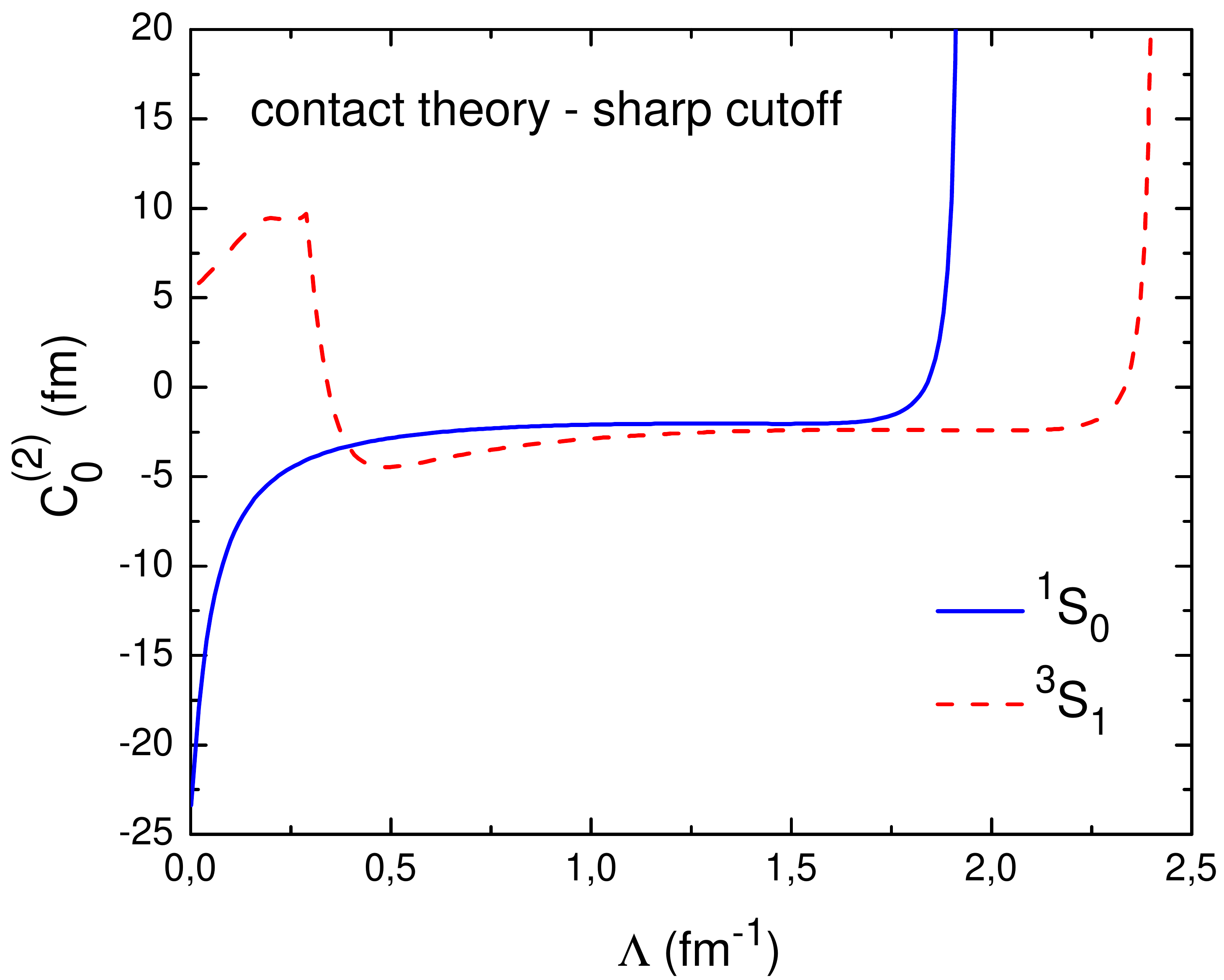}\hspace{0.2cm}
\includegraphics[width=5.5cm]{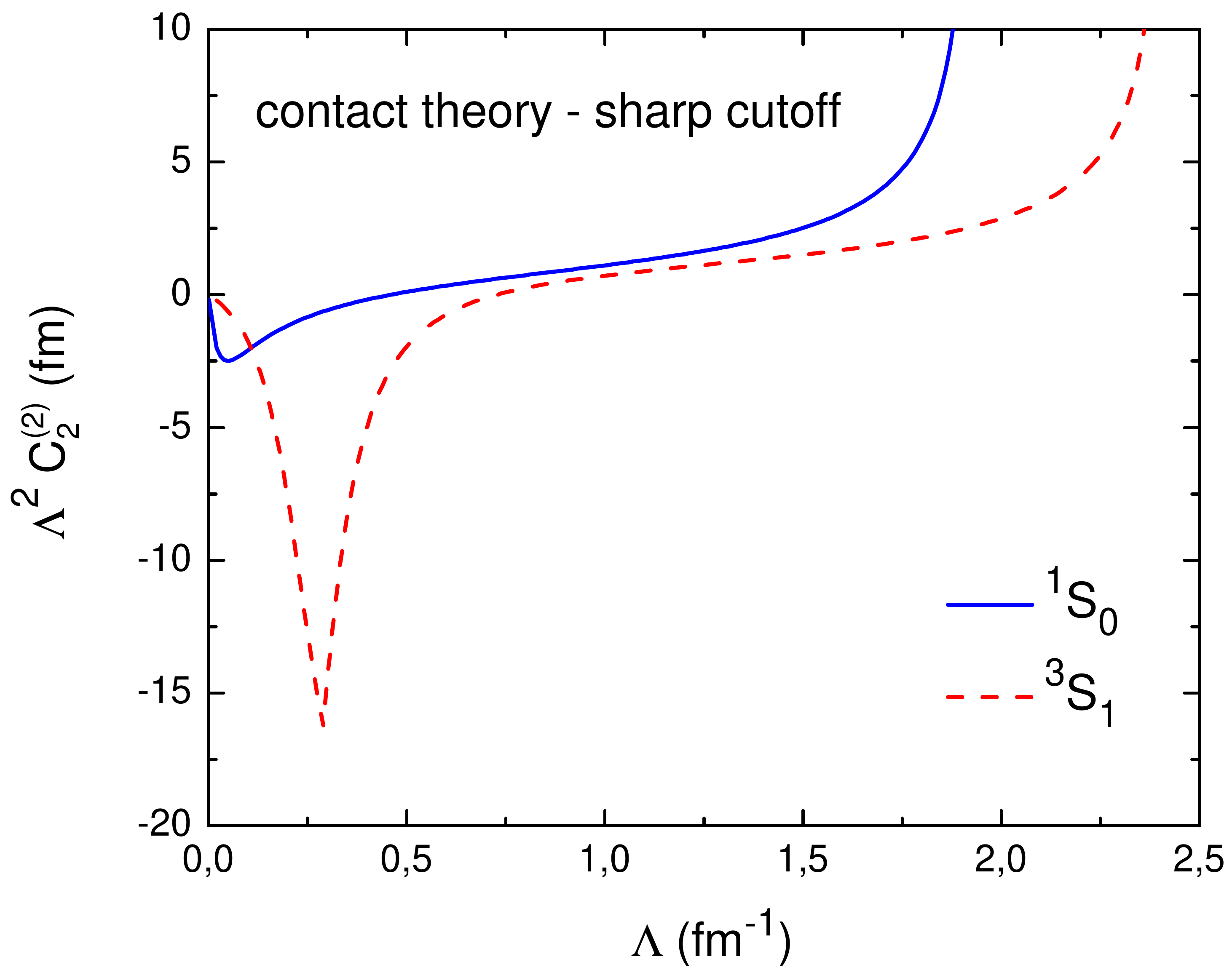}
\end{center}
\caption{$C_{0}^{(0)}$, $C_{2}^{(0)}$ and $C_{2}^{(2)}$ for the
  contact theory in the continuum regulated by a sharp momentum cutoff
  for the $^1S_0$ channel and the $^3S_1$ channel. The parameters are
  determined from the solution of the $LS$ equation for the on-shell
  $K$-matrix by fitting the ERE parameters}
\label{fig:2}
\end{figure*}

\section{Implicit Renormalization: Low cut-off evolution}

At low cut-offs $\Lambda$ we may approximate the hermitian effective
interaction by a polymomial,
\begin{eqnarray}
 V_\Lambda (p',p) 
&=& C_0 + C_2 (p^2 + p'^2) \nonumber \\ &+& 
C_4 (p^4+p'^4) + C_4' p^2 p'^ 2 + \dots  \, . 
\label{eq:pot-mom}
\end{eqnarray}
where $C_0,C_2, C_4 , C_4' , \dots$ are {\it real} coefficients
depending on $\Lambda$ to be determined. This corresponds to an
Effective Field Theory (EFT) with contact interactions only. We expect
Eq.~(\ref{eq:pot-mom}) to hold up to $p,p' \le \Lambda_{\rm
  ERE}$. Using the potential of Eq.~(\ref{eq:pot-mom}) the LS
Eq.~(\ref{eq:LS-mom-eff}) reduces to a system of algebraic equations
which solution is well known (see e.g.  Ref.~\cite{Entem:2007jg}). At
lowest leading order (LO) we just keep the leading term $C_0$ and get
\begin{eqnarray}
C_0 (\Lambda) = \frac{\alpha_0}{1-\frac{2\Lambda \alpha_0 }{\pi}} \, , 
\label{eq:C0-LO}
\end{eqnarray}
showing that $ \lim_{\Lambda \to 0 } V_\Lambda(0,0)= \alpha_0$. Going to
Next-to-leafing order (NLO) we obtain
\begin{eqnarray} 
-\frac1{\alpha_0 \Lambda } &=& \frac{4 \left(-2 c_2^2+90 \pi^4 +15 (3
  c_0+2 c_2) \pi ^2\right)}{9 \pi \left(c_2^2 -10 c_0 \pi ^2 \right)}
\, ,
\label{eq:r0-EFT}
\\ r_0 \Lambda &=& \frac{16 \left(c_2^2+12 \pi ^2 c_2+9
  \pi ^4\right)}{\pi \left(c_2+6 \pi ^2\right)^2} -\frac{12
  c_2 \left(c_2+12 \pi ^2\right)}{\left(c_2+6 \pi
  ^2\right)^2} \frac1{\alpha_0 \Lambda}  \, \nonumber \\
&+& \frac{3 c_2 \pi
  \left(c_2+12 \pi ^2\right)}{\left(c_2+6 \pi
  ^2\right)^2} \frac1{\alpha_0^2 \Lambda^2} \, , \nonumber 
\end{eqnarray} 
where $c_0 = 4 \pi \Lambda C_0$, $c_2 = 4 \pi \Lambda^3 C_2$. In the
second equation we have eliminated $C_0$ in terms of $\alpha_0$.  This
leads for any cut-off $\Lambda$ to the mapping $(\alpha_0,r_0) \to
(C_0, C_2) $. At this level of approximation there are two branches
and we choose the one consistent with the LO one for $\Lambda \to 0$,
see Eq.~(\ref{eq:C0-LO}) and Fig.~\ref{fig:2}. We will denote LO by
$C_{0}^{(0)}$ and NLO by $C_{0}^{(2)}$ and $C_{2}^{(2)}$. One should
note that in the case of the $^3S_1$ channel $C_{0}^{(0)}$ is singular
and the derivatives of $C_{2}^{(0)}$ and $C_{2}^{(2)}$ are
discontinuous at $\Lambda=\pi/2\alpha_0 \sim 0.3~{\rm fm^{-1}}$, which
is the momentum scale where the deuteron bound-state appears. The
strong resemblance of both $^1S_0$ and $^ 3S_1$ at the scales around
$\Lambda \sim 1~ {\rm fm}^{-1}$ is just a reminiscent of the SU(4)
Wigner symmetry for the two-nucleon
system~\cite{Arriola:2010hj,CalleCordon:2008cz,Timoteo:2011tt,Arriola:2013nja}.

One can in principle improve by including more terms beyond second
order in Eq.~(\ref{eq:pot-mom}).  The problem is that there are two
such terms $C_4$ and $C_4'$~\cite{Entem:2007jg} but there is only {\it
  one} low energy parameter in the ERE, $v_2$ in Eq.~(\ref{eq:ERE}).
This is so because scattering does not depend just on the on-shell
potential. Thus, the implicit renormalization is not unique beyond
NLO.  This is just a manifestation of the ambiguities of the inverse
scattering problem which can only be fixed after three or higher body
properties are taken into account~\footnote{Actually from a
  dimensional point of view the two-body operators with four
  derivatives are suppressed as compared to contact three body
  operators. The off-shellness of the two body problem can be
  equivalently be translated into some three-body
  properties~\cite{Amghar:1995av}.}.  Clearly, and even for the $C_0$
and $C_2$ coefficients, increasing $\Lambda$ values one starts seeing
more high energy details of the theory.

Even at NLO the question is how small must be the cut-off scale so
that Eq.~(\ref{eq:pot-mom}) works. There is a maximum value
$\Lambda_{\rm WB}$ for the cutoff scale $\Lambda$ above which one
cannot fix the strengths of the contact interactions $C_{2}^{(0)}$ and
$C_{2}^{(2)}$ by fitting the experimental values of both the
scattering length $\alpha_0$ and the effective range $r_0$ while
keeping the renormalized potential hermitian. This limit corresponds
to the Wigner causality bound realized as an off-shell unitarity
condition~\cite{Entem:2007jg,Szpigel:2010bj}. Indeed, for $\Lambda >
\Lambda_{\rm WB} \sim 1.9~\rm{fm}^{-1}$ in the case of the $^1S_0$
channel and $\Lambda > \Lambda_{\rm WB} \sim 2.4~\rm{fm}^{-1}$ in the
case of the $^3S_1$ channel, the parameters $C_{2}^{(0)}$ and
$C_{2}^{(2)}$ diverge before taking complex values and hence violating
the hermiticity of the effective potential in Eq.~(\ref{eq:pot-mom}).

\begin{figure*}[ht]
\begin{center}
\includegraphics[width=4cm]{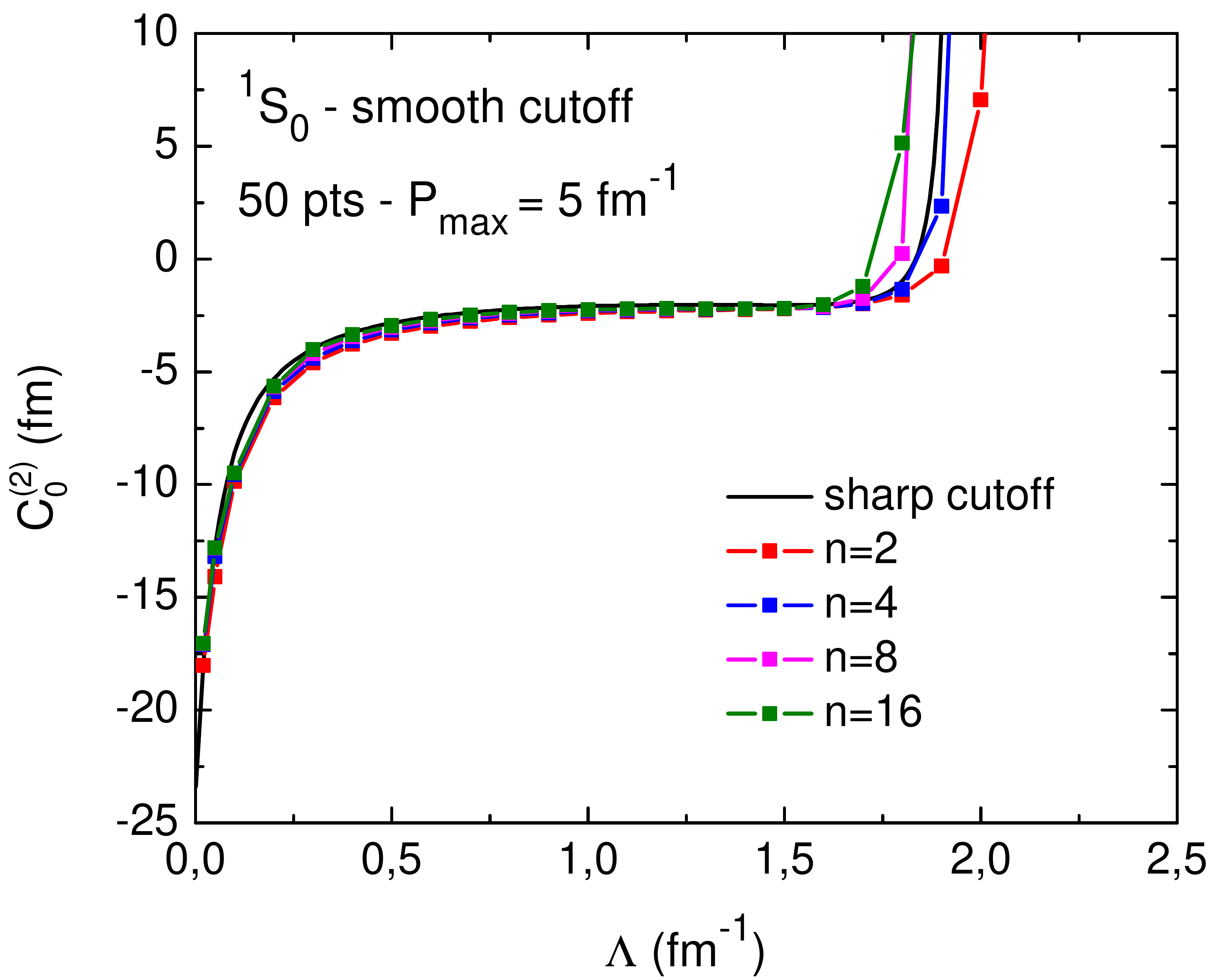}
\includegraphics[width=4cm]{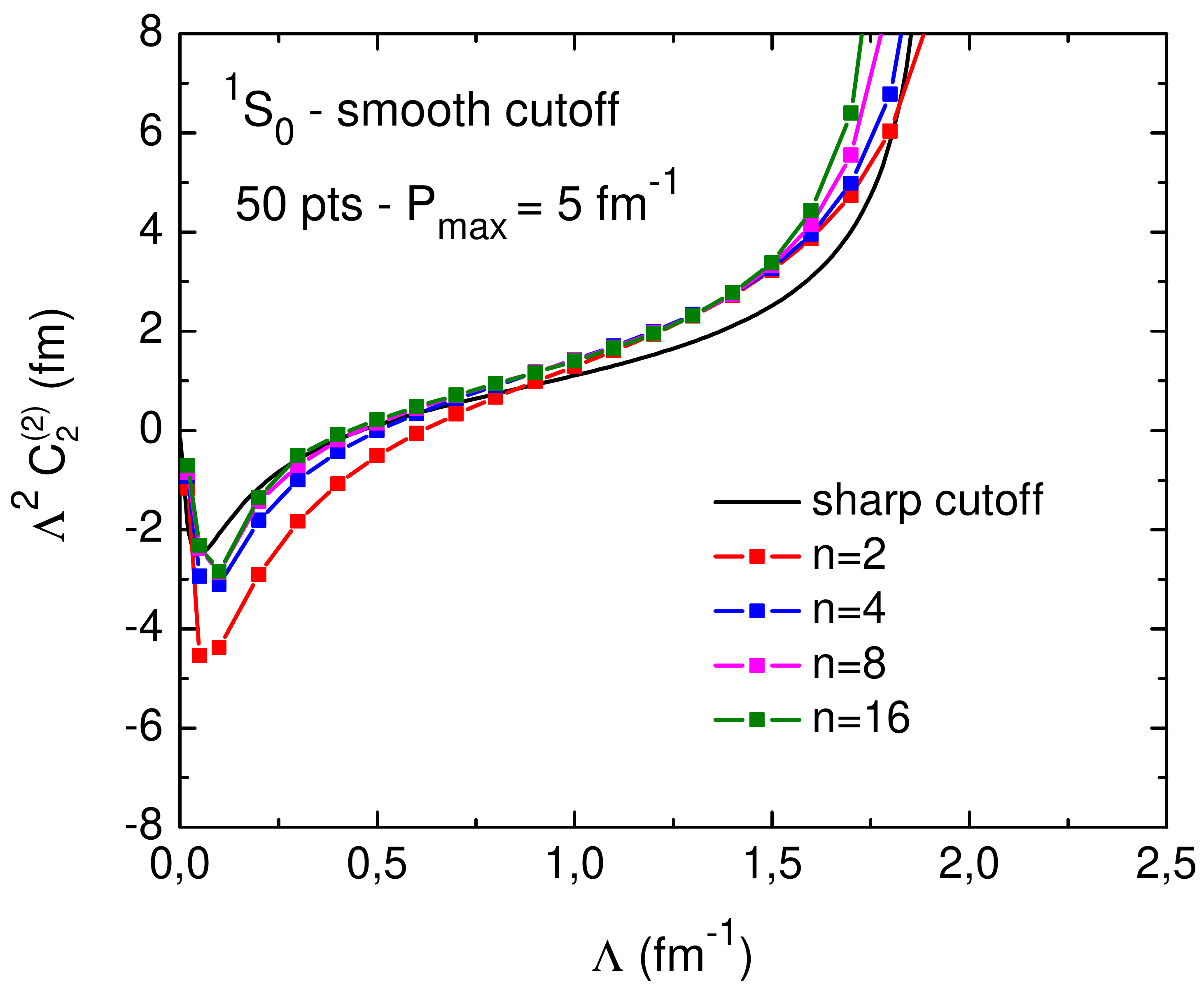}
\includegraphics[width=4cm]{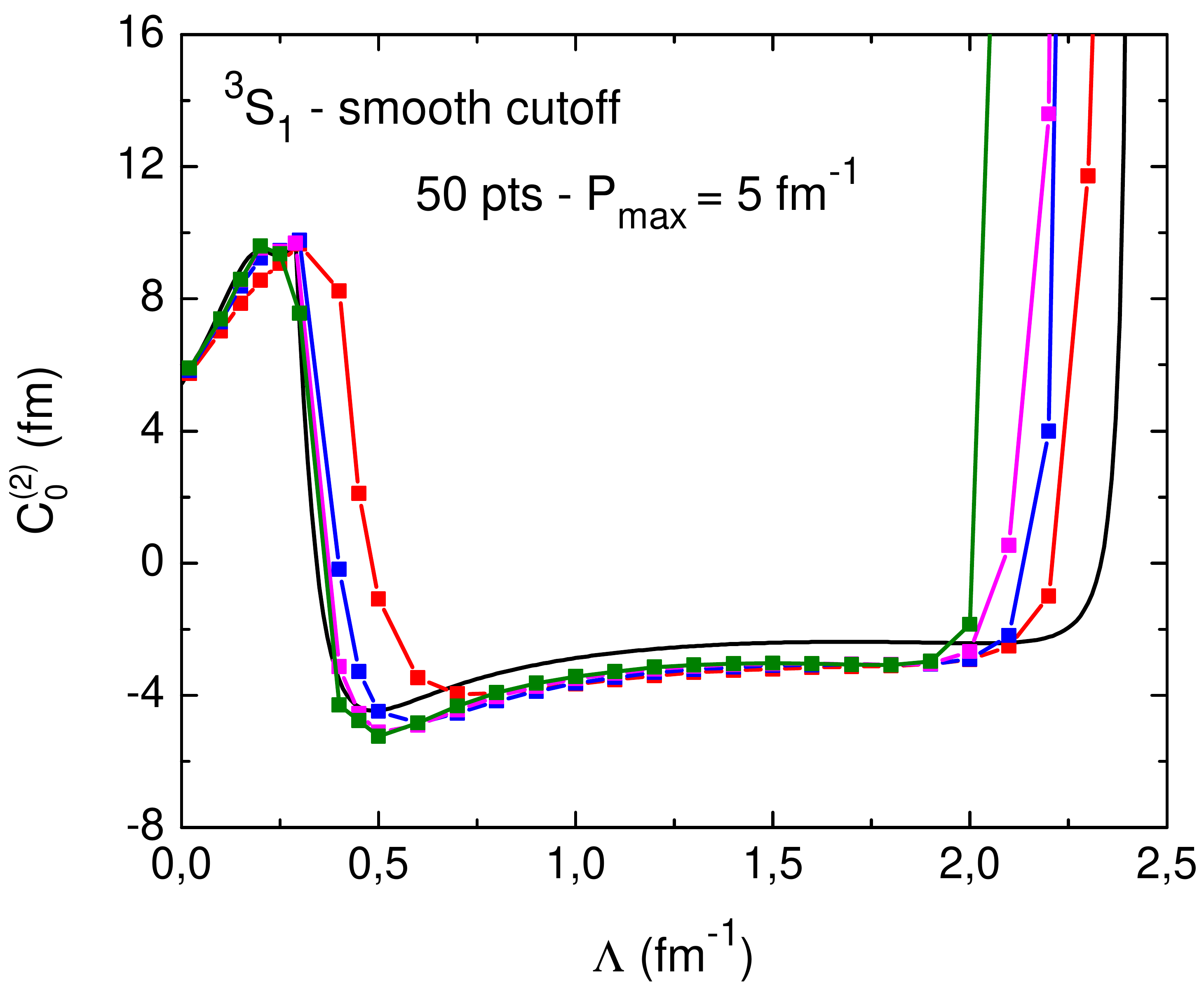}
\includegraphics[width=4cm]{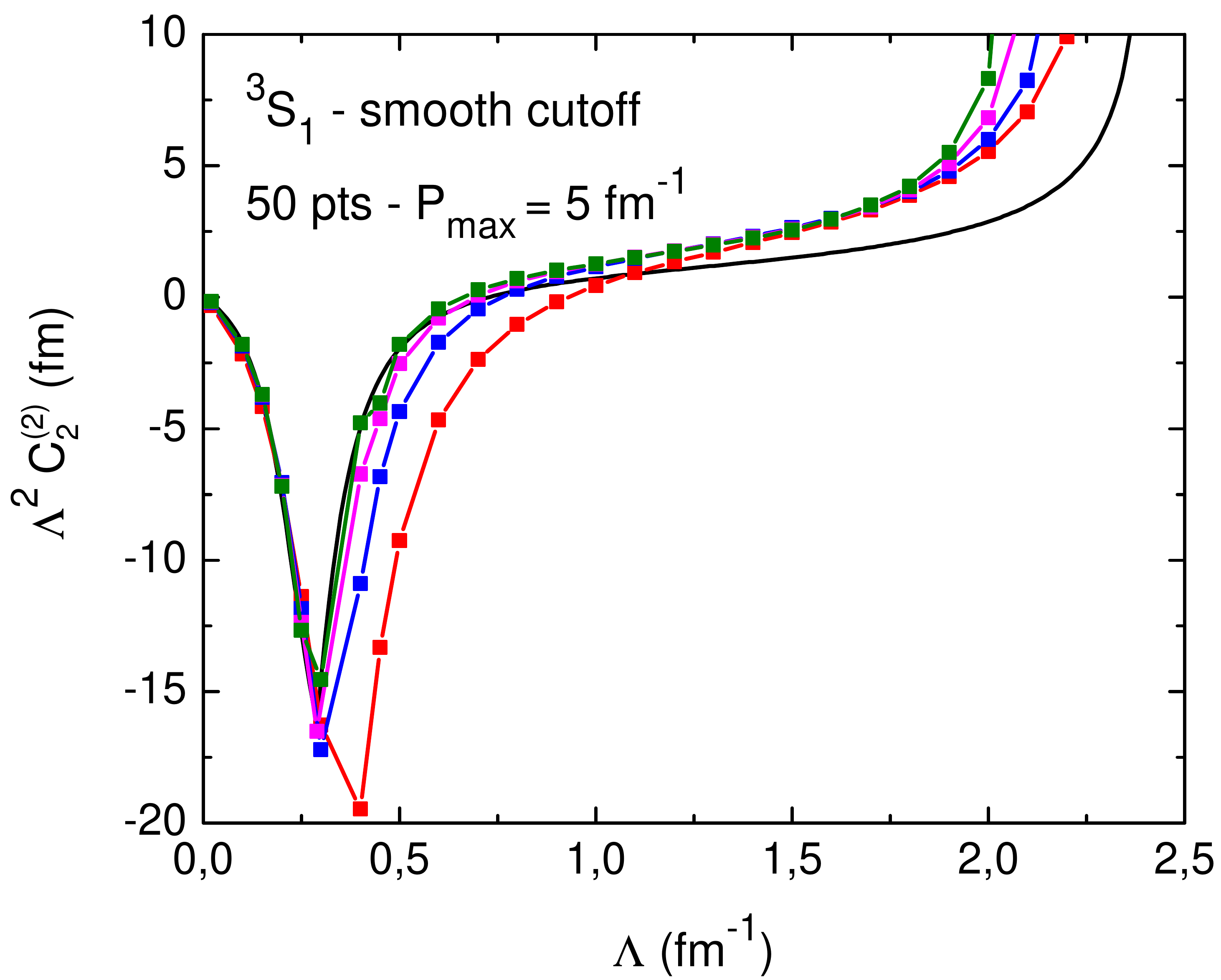}\\\vspace{0.1cm}
\includegraphics[width=4cm]{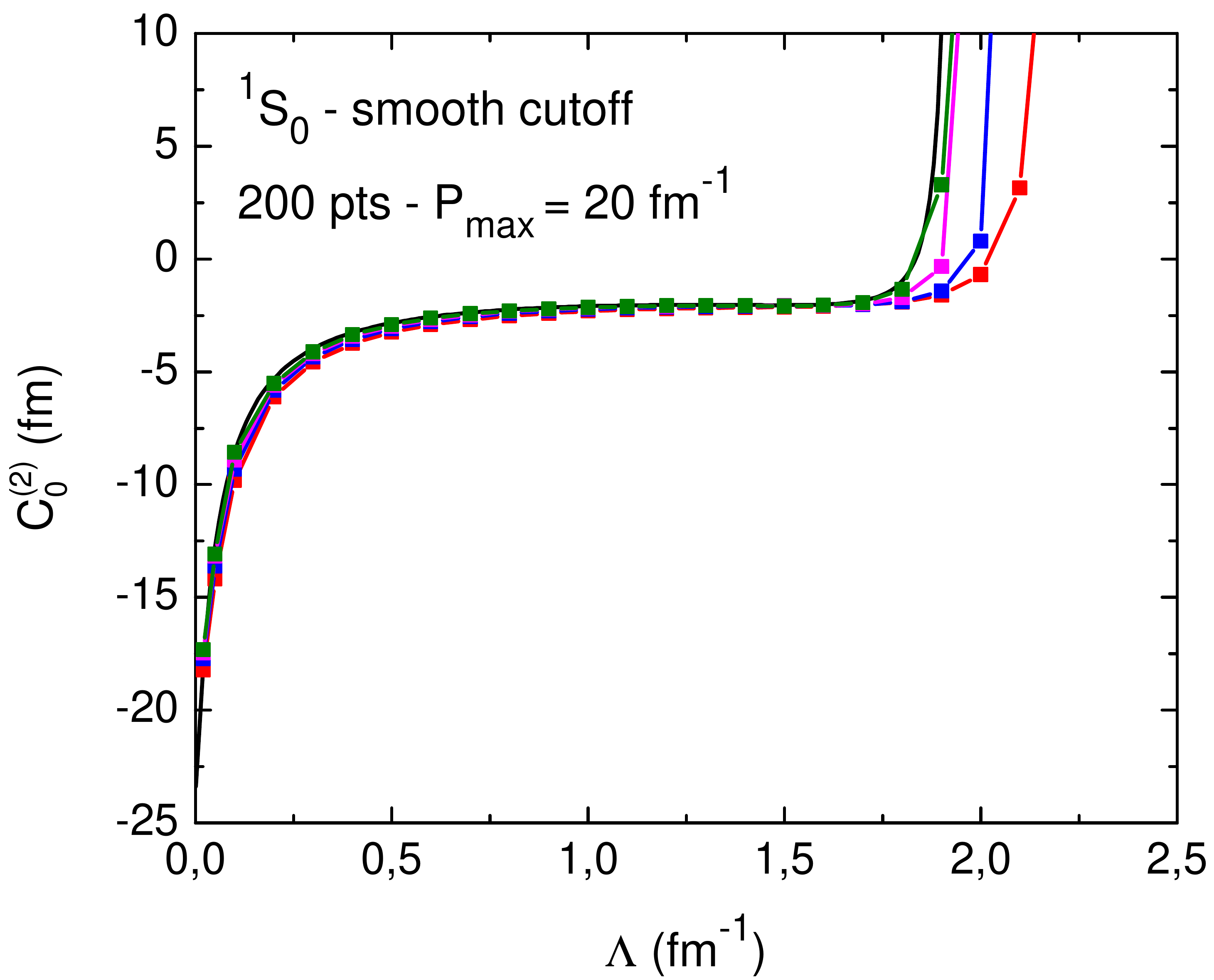}
\includegraphics[width=4cm]{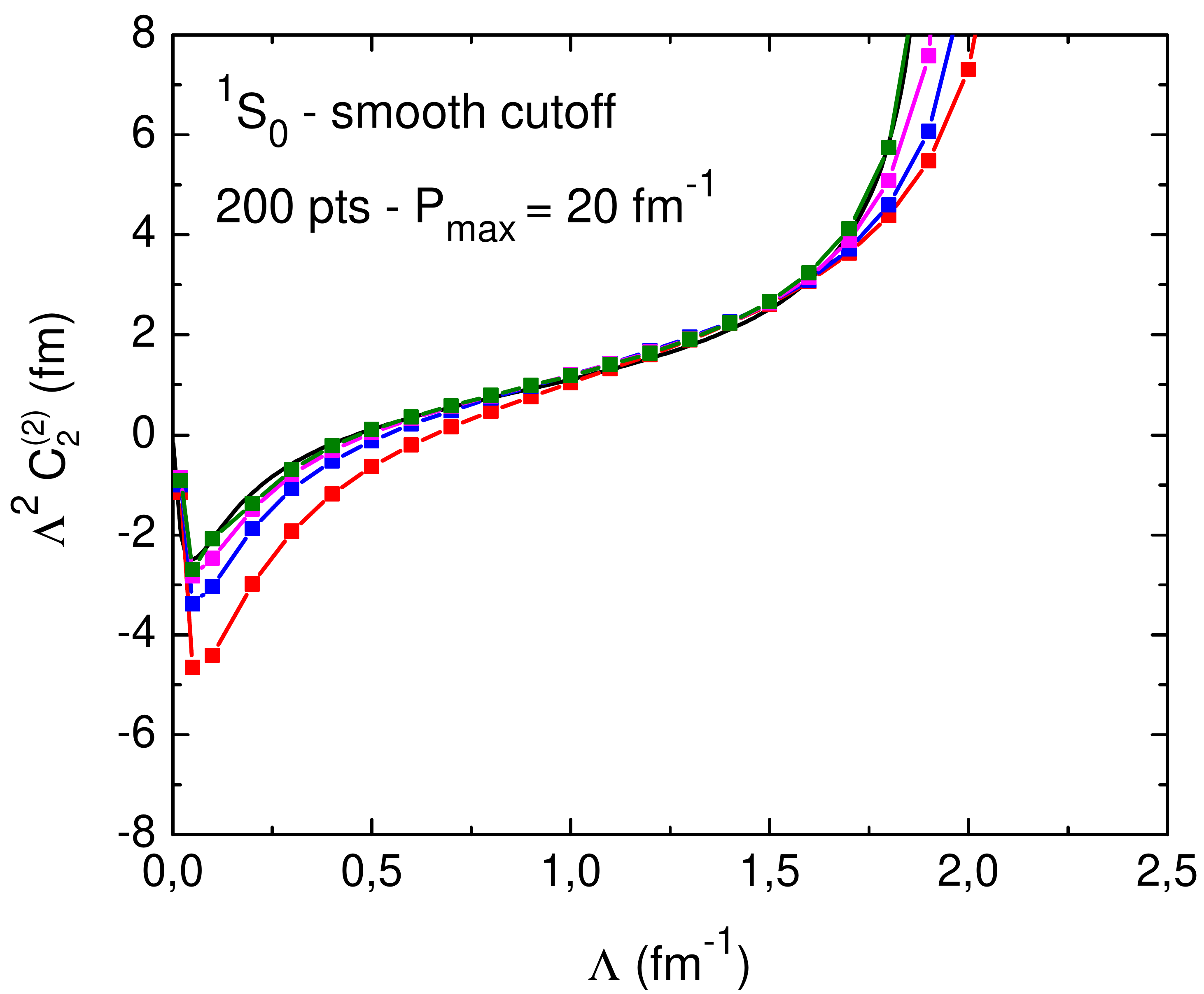}
\includegraphics[width=4cm]{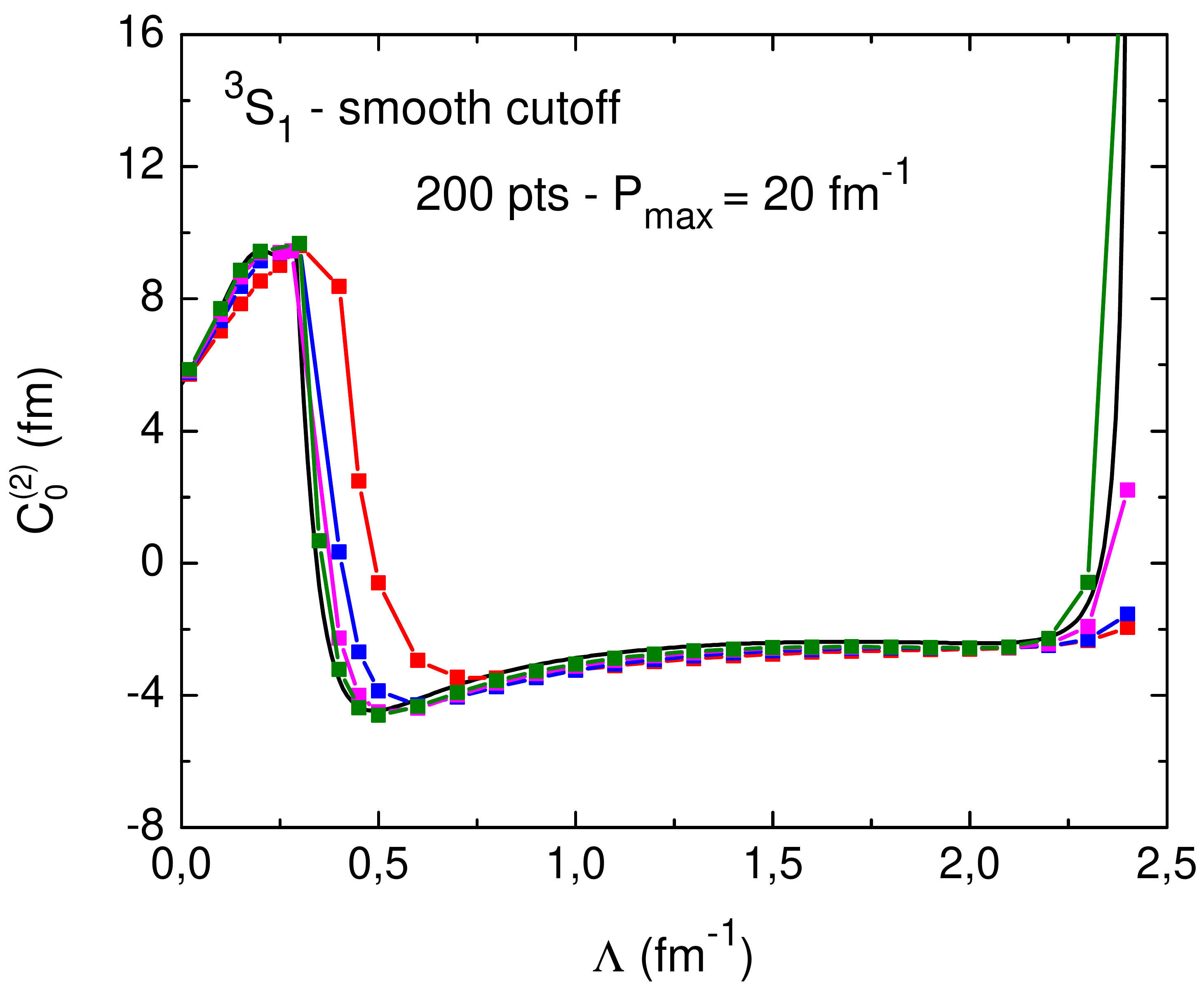}
\includegraphics[width=4cm]{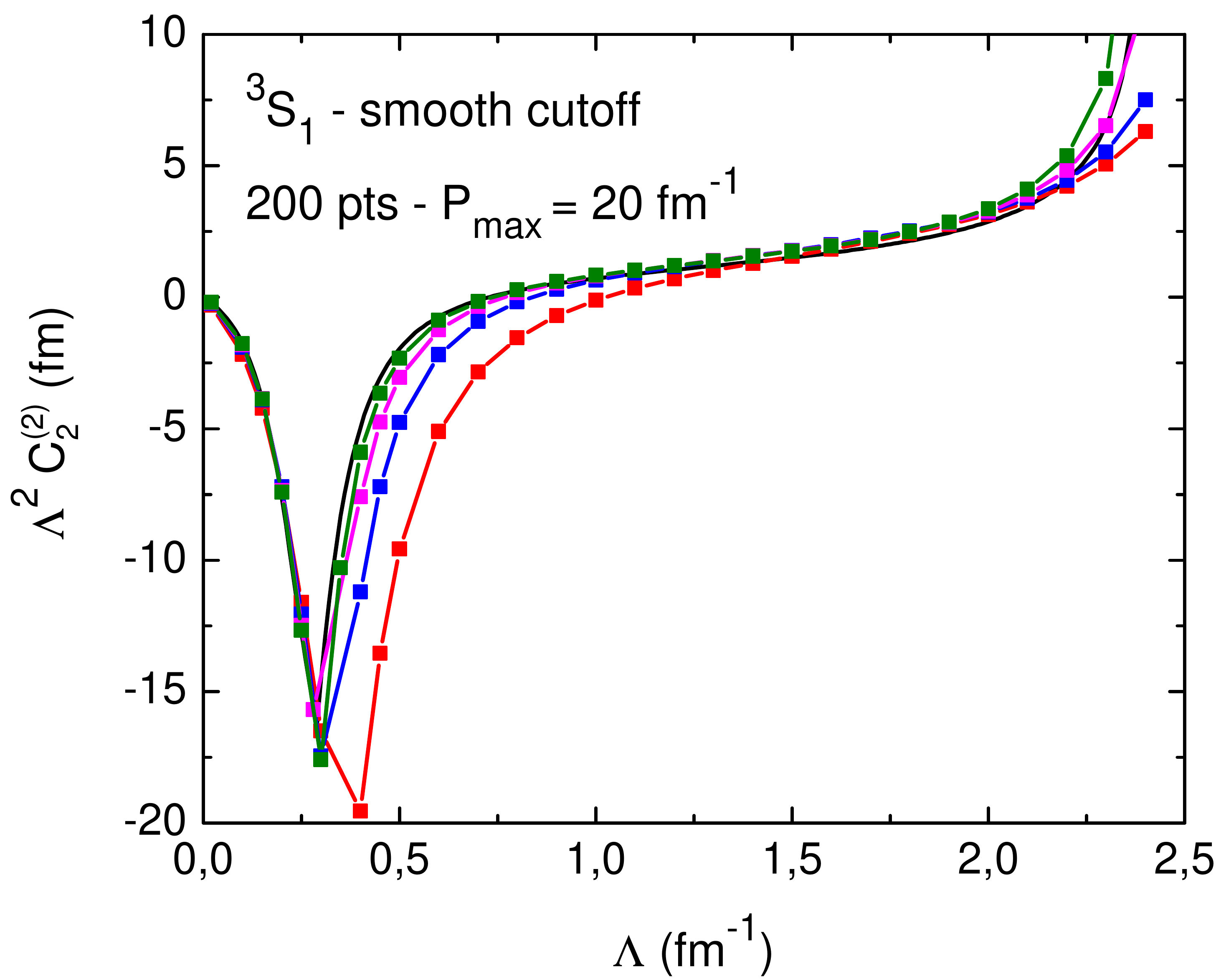}\\\vspace{0.1cm}
\includegraphics[width=4cm]{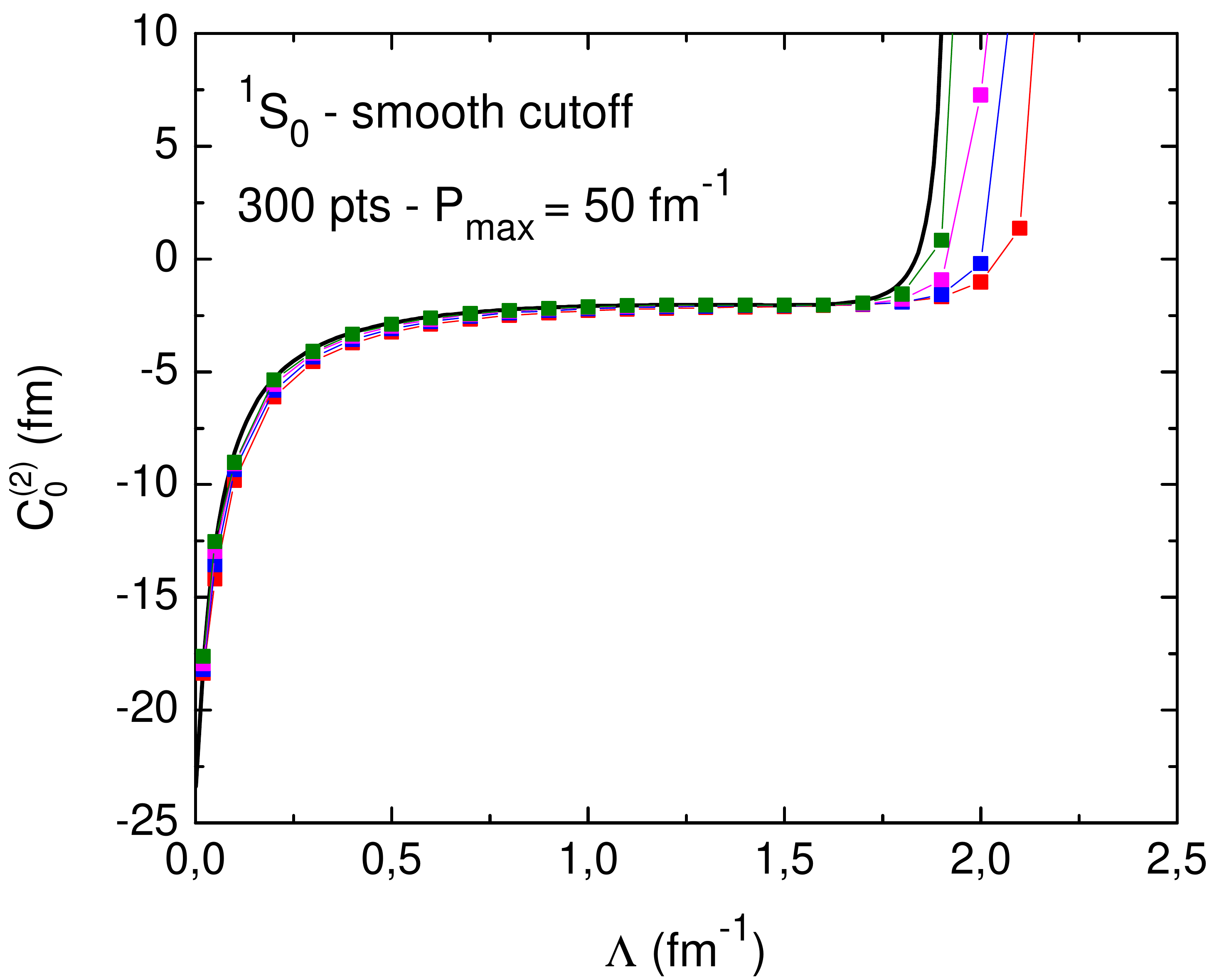}
\includegraphics[width=4cm]{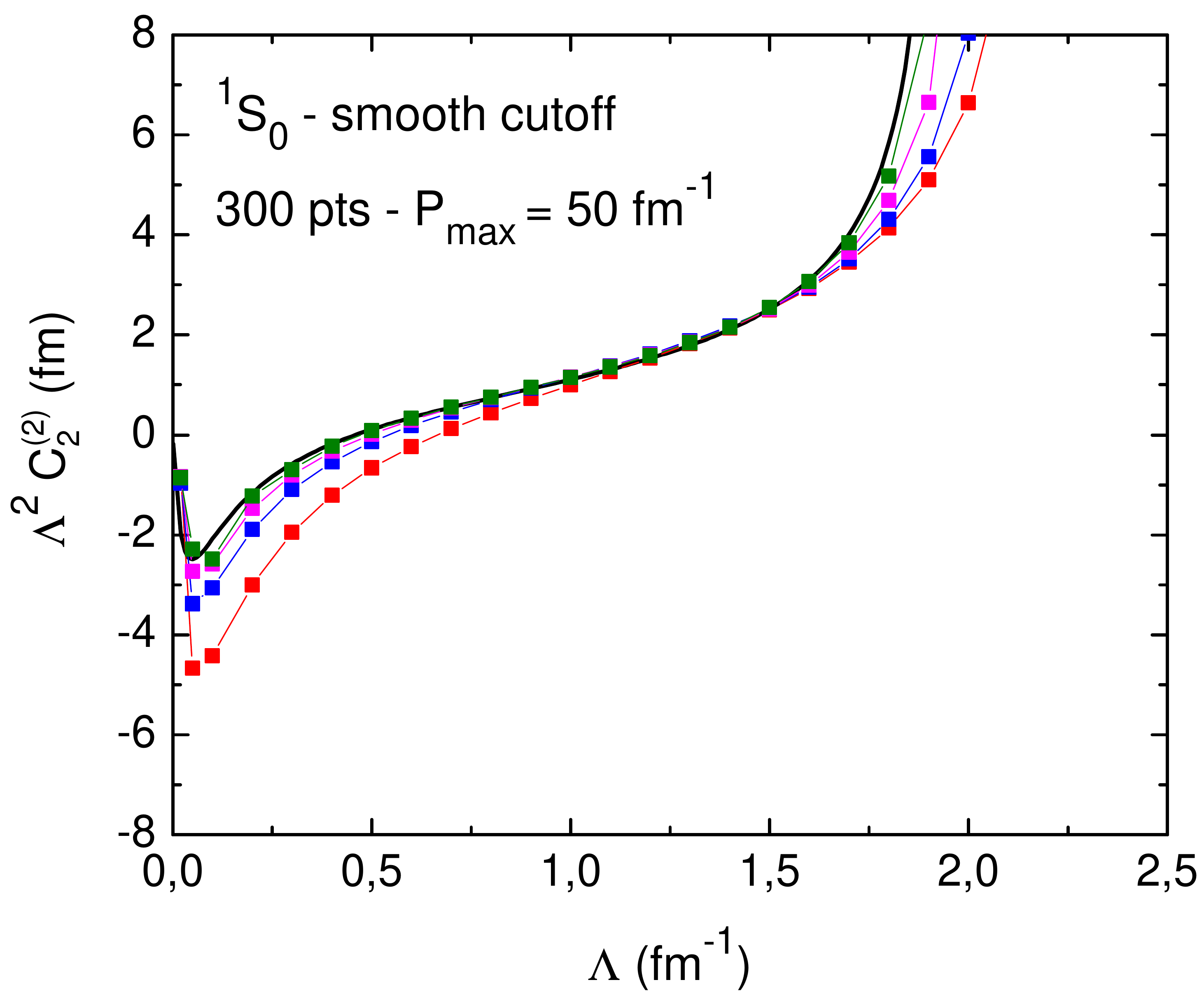}
\includegraphics[width=4cm]{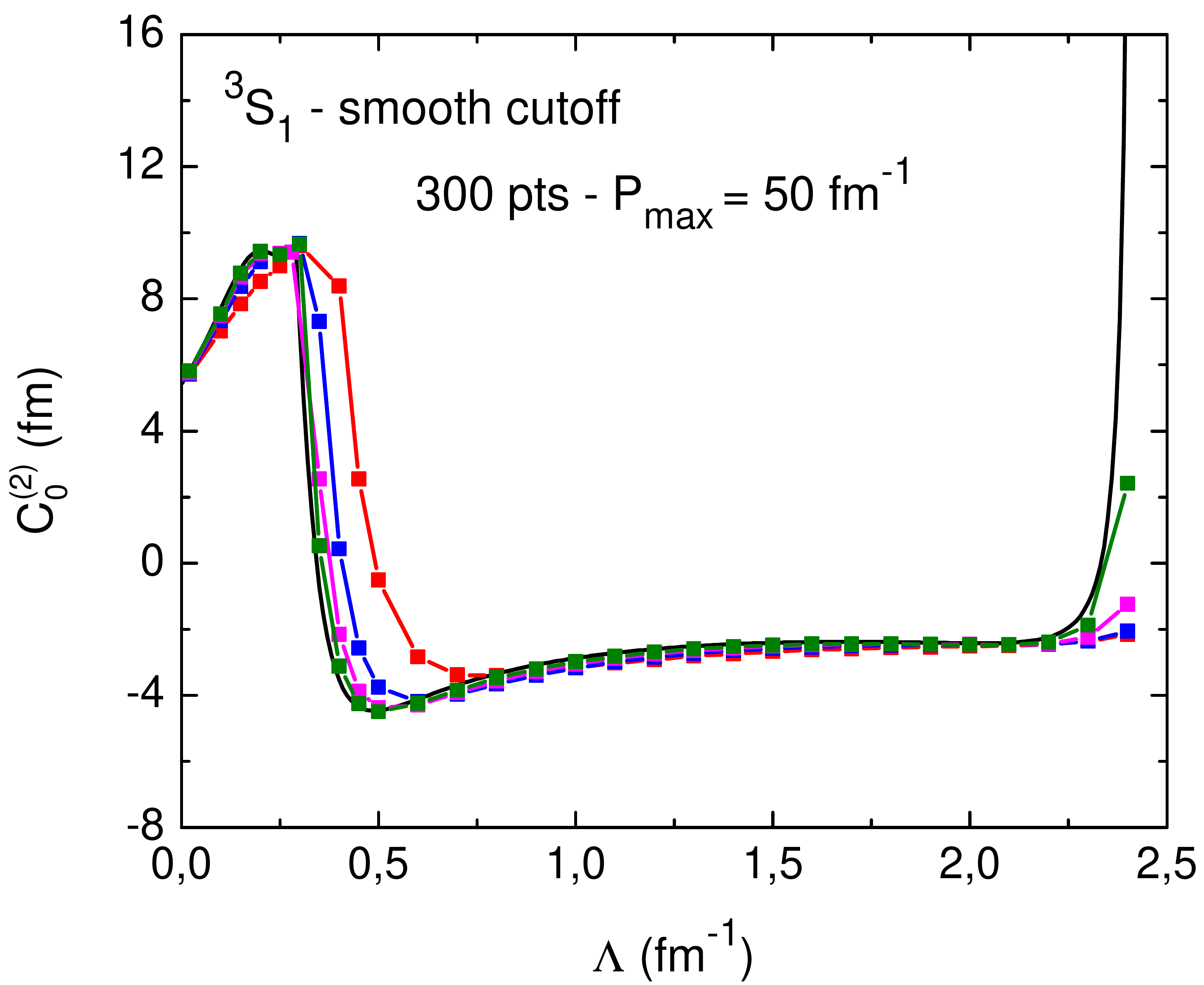}
\includegraphics[width=4cm]{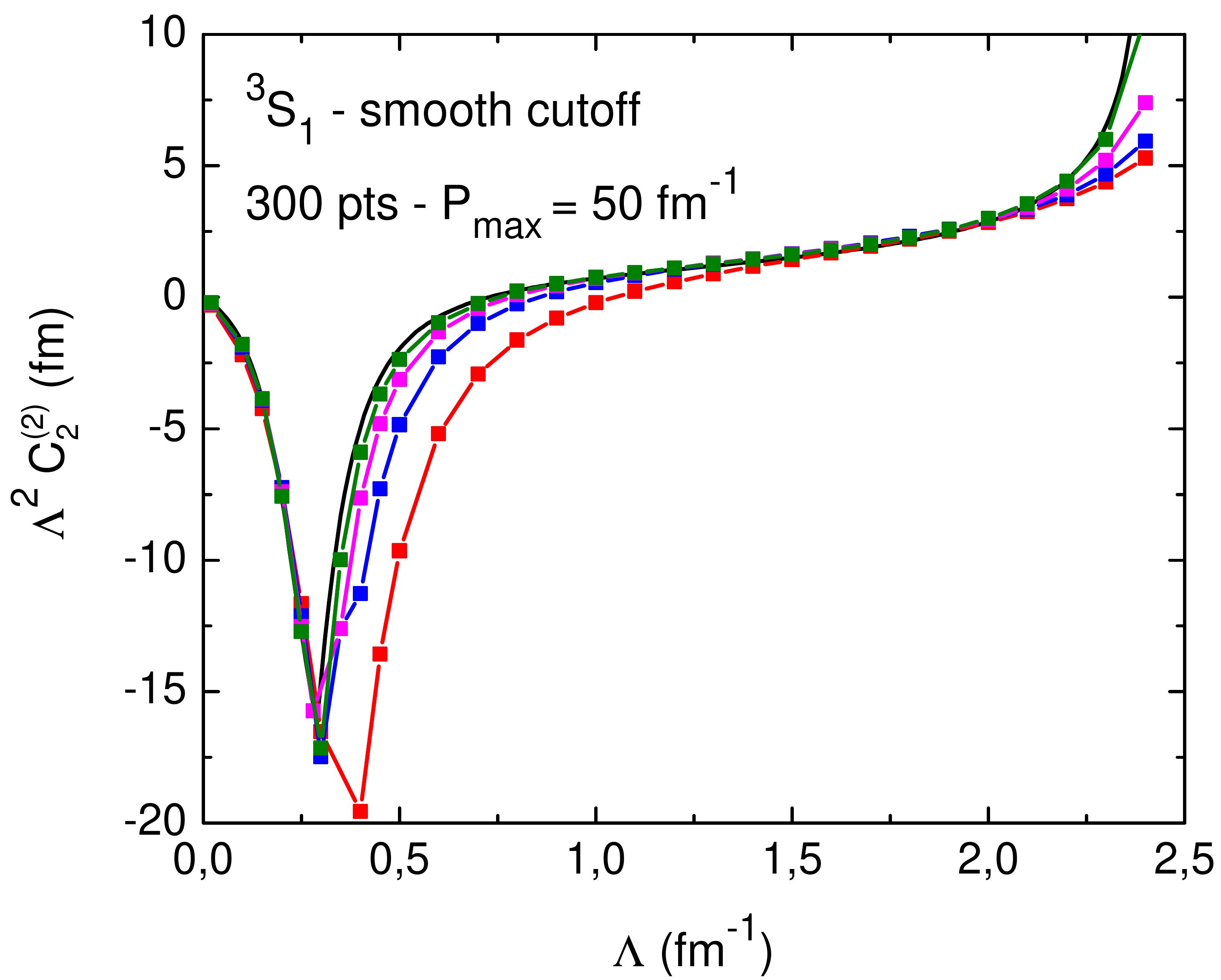}
\end{center}
\caption{$C_{0}^{(2)}$ and $C_{2}^{(2)}$ for the contact theory on a
  grid regulated by a smooth exponential momentum cutoff for the
  $^1S_0$ channel and the $^3S_1$ channel $NN$ potentials at NLO. For
  comparison, we also show the corresponding $C_{0}^{(2)}$ and
  $C_{2}^{(2)}$ for the contact theory in the continuum regulated by a
  sharp momentum cutoff. In both cases the parameters are determined
  from the solution of the $LS$ equation for the on-shell $K$-matrix
  by fitting the ERE parameters.}
\label{fig:grids-50-300}
\end{figure*}

\begin{figure*}[ht]
\begin{center}
\includegraphics[width=5.5cm]{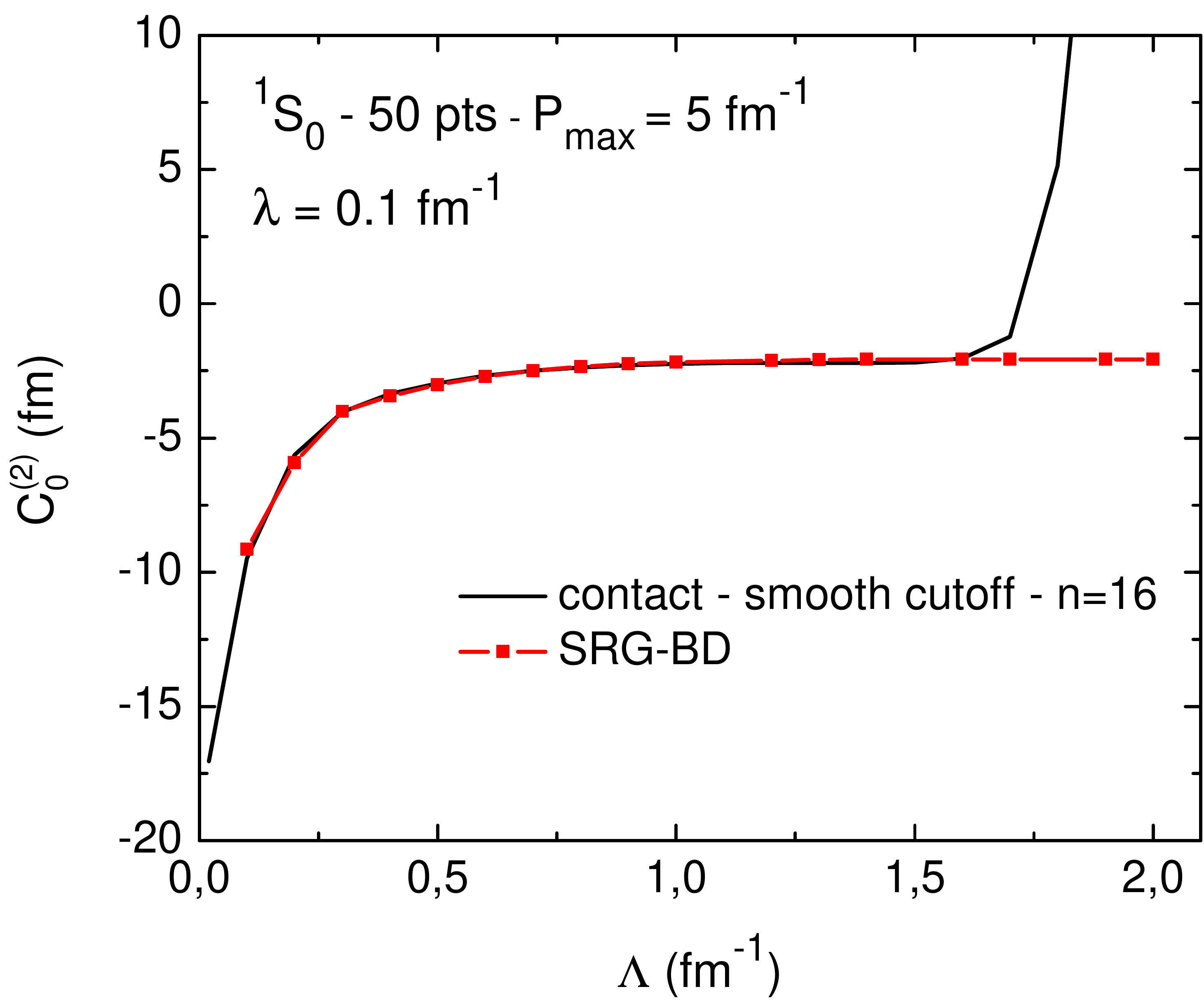}\hspace{0.5cm}
\includegraphics[width=5.5cm]{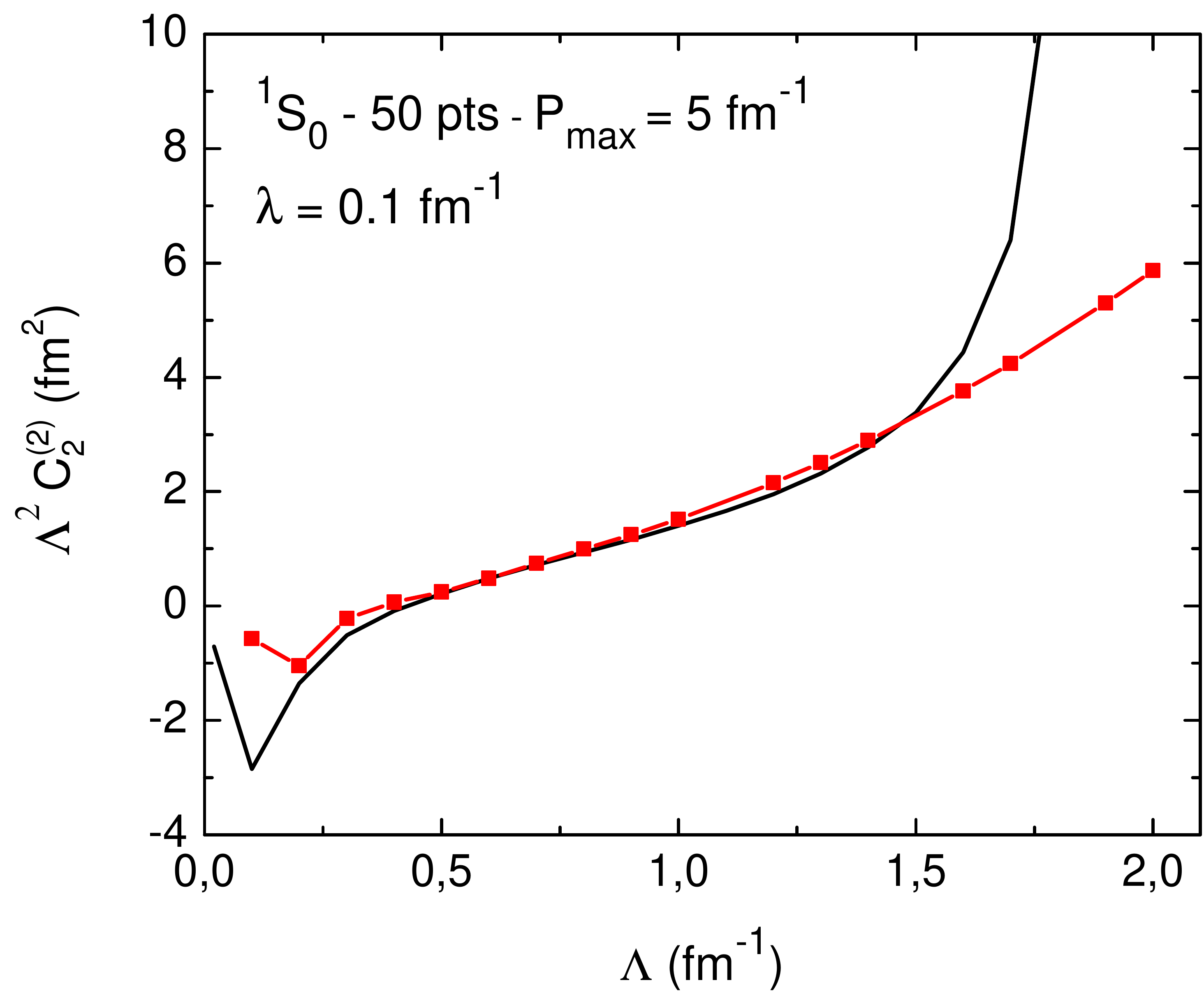} \\\vspace{0.1cm}
\includegraphics[width=5.5cm]{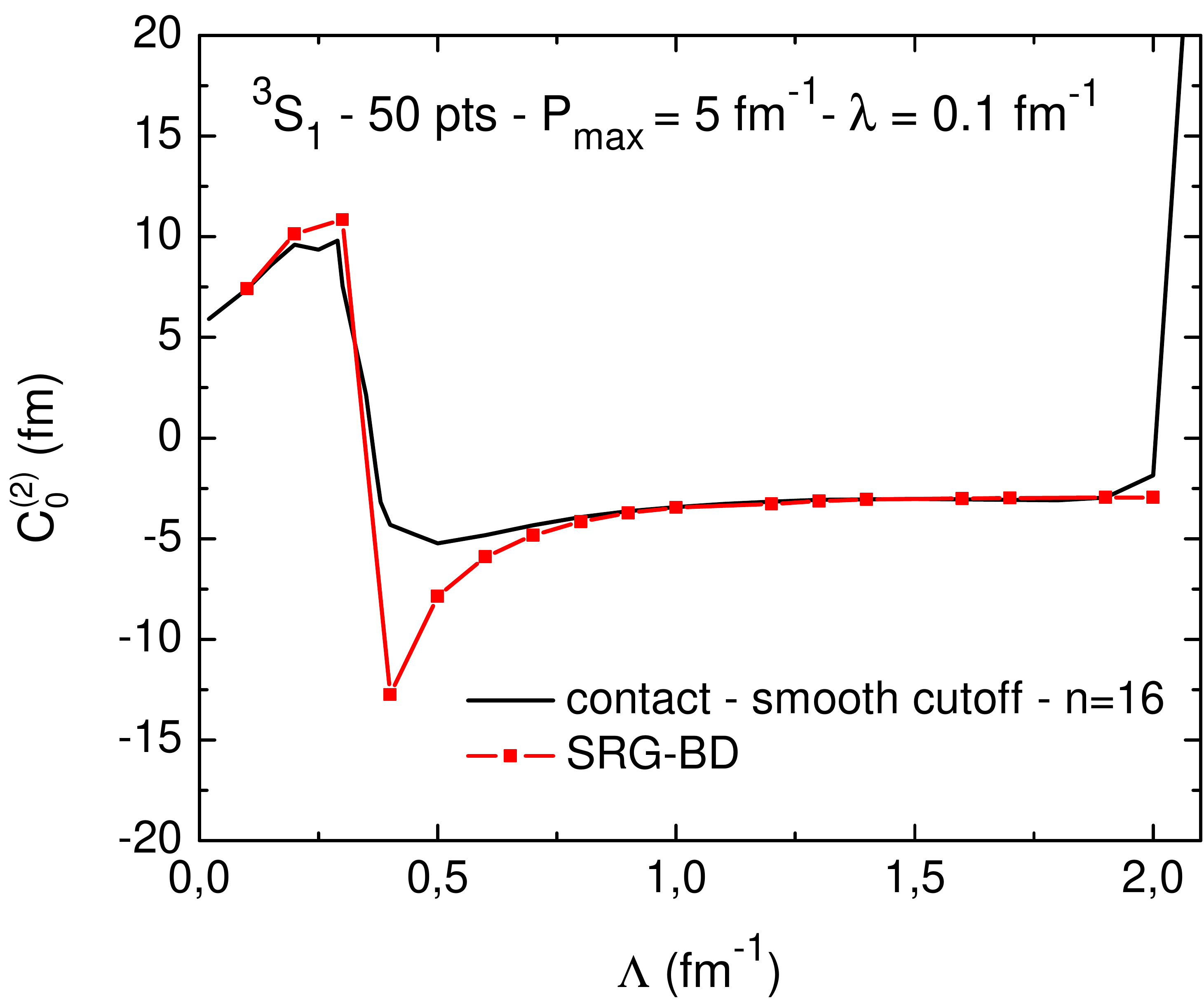}\hspace{0.5cm}
\includegraphics[width=5.5cm]{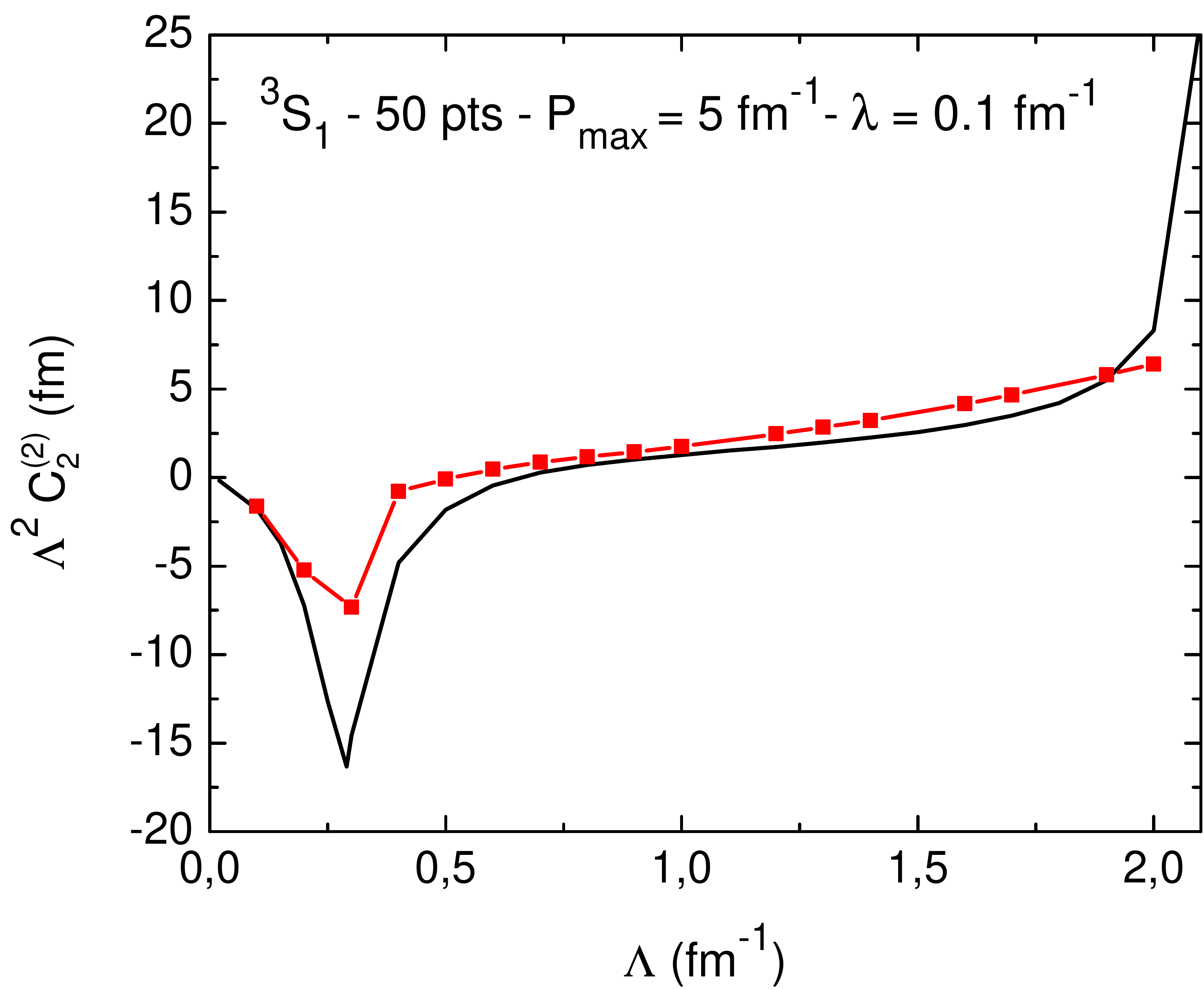} 
\end{center}
\caption{$C_0$ and $C_2$ as a function of the block-diagonal cutoff
  $\Lambda$ extracted from the $^1S_0$ and $^3S_1$ channels gaussian
  potential on a grid (with $N=50$ gauss points and $P_{\rm
    max}=5~{\rm fm}^{-1}$) evolved through the SRG transformation with
  the block-diagonal generator for the lowest SRG cutoff
  $\lambda=0.1{\rm fm}^{-1}$. For comparison, we also show $C_0$ and $C_2$
  for the $^1S_0$ and $^3S_1$ channels contact theory potential at
  NLO (on the same grid) regulated by a smooth exponential momentum
  cutoff with $n=16$.}
\label{fig:18}
\end{figure*}

\section{Numerical results}

The Block-Diagonal-SRG equations, Eq.(\ref{eq:SRG}), have to be solved
numerically on a momentum grid with $N$-points yielding $4 \times N^2$
non-linear first order coupled differential equations. Furthermore, an
auxiliary numerical cut-off $P_{\rm max}= N \Delta p$ must also be
introduced. It is interesting to test the space dimensions needed to
solve the contact theory close to the continuum. This is shown in
Fig.~\ref{fig:grids-50-300} where one sees that large $N$
is needed to reproduce the continuum limit. We will set $N=50$ and
$P_{\rm max}= 5~ {\rm fm}^{-1}$ to our SRG calculations, solve the
system of $4\times N^2$ non-linear first-order coupled differential
equations by using an adaptative fifth-order Runge-Kutta algorithm as
in Ref.~\cite{Szpigel:2010bj} and compare the results to the contact
interaction with the same $N$ and $P_{\rm max}$. We check unitarity by
comparing phase-shifts along the $\lambda$ evolution, $\delta_\lambda
(p)=\delta(p)$. The sharp momentum projectors in
Eq.~(\ref{eq:PQ-sharp}) may be regularized as smooth
projectors~\cite{Bogner:2006vp} ($Q \equiv 1-P$)
\begin{eqnarray}
P = \Theta(\Lambda-p) = \lim_{n \to \infty} e^{-(p/\Lambda)^n} \, , 
\end{eqnarray}
and we will take the values $n=2,4,8,16$ to check convergence.
We want to compare the running of the coefficients $C_0$ and $C_2$ with
the cut-off $\Lambda$ in the contact theory potential at NLO to the
running of the corresponding coefficients ${\tilde C_0}$ and ${\tilde
  C_2}$ with the $V_{\rm low~k}$ cutoff ($\equiv \Lambda$) extracted from a
  polynomial fit of the BD-SRG-evolved gaussian potential, 
\begin{equation}
V_{\lambda,\Lambda}(p,p') = {\tilde C_{0}} + {\tilde
  C_{2}}~(p^2+{p'}^2) + \cdots\; .
\end{equation}
\noindent
The parameters $C$ and $L$ in the initial gaussian potential
($\lambda, \Lambda \rightarrow \infty$), defined by
Eq.~(\ref{gaupot}), and the coefficients $C_{0}$ and $C_{2}$ in the
contact theory potential at NLO are determined from the solution of
the LS equation for the $K$-matrix on the finite momentum grid by
fitting the experimental values of the scattering length $\alpha_0$
and the effective range $r_0$.  The coefficients ${\tilde C_{0}}$ and
${\tilde C_{2}}$ are determined by fitting the diagonal
matrix-elements of the BD-SRG-evolved potential for the lowest momenta
with the polynomial form and the finite momentum
grid~\footnote{Actually, looking for a fiducial region to extract
  $C_0$ and $C_2$ in the explicit method when $\Lambda \ge
  \Lambda_{\rm ERE}$ requires allowing for higher order polynomial
  contributions.}.

In Fig.~\ref{fig:18} we show the results for ${\tilde C_{0}}$ and
$\Lambda^2 {\tilde C_{2}}$ extracted from the $^1S_0$ channel and the
$^3S_1$ channel BD-SRG-evolved gaussian potentials on a grid (with
$N=50$ gauss points and $P_{\rm max}=5~{\rm fm}^{-1}$) and down to the
lowest SRG cutoff $\lambda=0.1~{\rm fm}^{-1}$, compared to $C_0$ and
$\Lambda^2 C_2$ obtained for the contact theory potential at NLO (on
the same grid) regulated by a smooth exponential momentum cutoff with
sharpness parameter $n=16$. As we see, there is a remarkably good
agreement between the coefficients extracted from the BD-SRG-evolved
potential and those obtained for the contact theory in the limit
$\lambda \rightarrow 0$.

It is important to point out that the agreement between the running of
the coefficients $C_{0}$ and $C_{2}$ in the contact theory potential
and the running of the coefficients ${\tilde C_{0}}$ and ${\tilde
  C_{2}}$ extracted from the BD-SRG-evolved gaussian
potential as the similarity cutoff $\lambda$ decreases below $\Lambda$
can be traced to the decoupling between the $P$-space and the
$Q$-space, which follows a similar pattern. Thus, in the limit
$\lambda \rightarrow 0$ we expect to achieve a high degree of
agreement for cutoffs $\Lambda$ up to $\Lambda_{\rm WB}$ determined by
the Wigner bound for the contact theory.

The overlapp between the discretized explicit and implicit numerical
solutions is verified in a wide range of cut-offs $\Lambda$. If
continuum accuracy was to be judged from the slow convergence pattern
of Fig.~\ref{fig:grids-50-300}, the equivalent BD-SRG calculations
would be out of question. Thus, the continuum limit $\Delta p \to 0$
is better and more simply represented by the implicit approach.

For the $^1S_0$ and $^3S_1$ neutron-proton scattering states this
range is within $ 0.5~ {\rm fm}^{-1} \le \Lambda \le 1.5~ {\rm
  fm}^{-1} $. This is a welcome feature, since it suggests that the
bulk of the effective interaction and its scale dependence can
directly be extracted from low energy NN data, as done in
Ref.~\cite{Arriola:2010hj}, where the Skyrme force parameters
deducible solely from the NN interaction in S- and P-waves were
determined.

It is interesting to determine the role played by OPE in the implicit
method in the cut-off range around $\Lambda \sim m_\pi $ but below
pion production threshold $\Lambda \lesssim \sqrt{m_\pi M_N} $ as OPE
is an indispensable ingredient of realistic bare interactions (see
e.g. Ref.~\cite{Harada:2010ba}). According to the recent Partial Wave
Analysis of Ref.~\cite{Perez:2013mwa} of about 8000 pp and np data,
OPE is the only needed contribution for $r> 3~{\rm fm}$. If one
separates the initial condition as $V_{\lambda=\infty}= V_{r \le 3
  {\rm fm}} + V_{1\pi, ~r \ge 3 {\rm fm}} \equiv V_S + V_L $ one has
$V_L \equiv V_{1\pi,~ r \ge 3 {\rm fm}} \ll V_S \equiv V_{r \le 3 {\rm
    fm}} $ and one can attempt a perturbative expansion of the
block-SRG evolved potential.  Thus, evolving the full $V$ and {\it
  just} $V_S$ we find from Eq.~(\ref{eq:SRG}) that $ P V_{\rm low k} P
= P V_{S,\rm low k} P + P V_L P + {\cal O} ( V_L^2)$. Using the
$\delta$-shell representation of Ref.~\cite{Perez:2013mwa} we get that
the accuracy of this perturbation theory in the $^1S_0$ channel is
indeed small; ${\cal O} ( V_L^2) \le 10^{-2} {\rm fm}$ for $2.1 \ge
\Lambda \ge 0.5 {\rm fm}^{-1}$ at $p,p' \le \Lambda$. This suggests
that the unevolved (and $\Lambda$-independent) long distance OPE piece
($ r \ge 3 {\rm fm}$) remains small after evolution and this
contribution can be treated perturbatively. A more complete analysis
of this important issue will be presented elsewhere.


The block diagonal SRG reduces the model space but also induces a
truncation error for decreasing $\Lambda$. The SRG evolves the bare
hamiltonian to a lower similarity cutoff $\lambda$. Due to the
unitarity, the SRG evolution to a lower similarity cutoff $\lambda$
preserves the EFT truncation errors for a given $V_{\rm low-k}$ cutoff
$\Lambda$.

We have shown with a simple example the situation with S-waves. Higher
partial waves, such as P-waves are in better shape. The scale
saturation displayed by the implicit method in
Ref.~\cite{Arriola:2010hj} becomes more pronounced in this case (see
Eq.(12) in that paper). It remains to be seen what is the actual
situation when the implicit method is applied to interactions
describing NN scattering data to higher energies. Work along these
lines is in progress.

\section{Conclusions}

While the effective interaction idea is very appealing there is no
unique way to define it; its definition depends on {\it how} are the
high and low energies separated and {\it what} is the relevant energy
scale cut-off. Within a given scheme, however, the cut-off scale
dependence of effective interactions representing a given bare
interaction in a model space can be carried out in two complementary
ways: either explicitly from a Block-Diagonal SRG transformation or
implicitly by using scattering data and renormalization
conditions. Although the complementarity of both explicit and implicit
views of the renormalization procedure is often invoked on general
grounds, we note that it is seldomly tested within the present context
of nuclear effective interactions.  As we have shown such a test
requires to pin down the numerics in a finite momentum grid with
sufficient accuracy making the explicit BD-SRG method computationally
expensive and impractical. At low energies effective interactions and
their scale dependence are just given by counterterms evaluated for
finite cut-offs. We find a remarkably wide range of cut-offs where
this complementarity holds in a model independent way. This suggests
that the implicit renormalization approach may be a simpler, more
accurate and direct method to determine the effective interaction than
the explicit and traditional method based on numerically integrating
the operator SRG flow equations using as initial condition a
phenomenological bare interaction fitted to NN scattering data below
pion production threshold. Another related issue is the role played by
three--, four-body etc. properties in the definition of two body
effective interactions as this becomes necessary for a truly model
independent formulation of effective interactions.

E.R.A. was supported by Spanish DGI (grant FIS2011-24149) and Junta de
Andalucia (grant FQM225). S.S.  was supported by Instituto
Presbiteriano Mackenzie and FAPESP. V.S.T.  would like to thank
FAEPEX, FAPESP and CNPq for financial support.  Computational power
provided by FAPESP grant 2011/18211-2.


\begin{thebibliography}{32}
\expandafter\ifx\csname natexlab\endcsname\relax\def\natexlab#1{#1}\fi
\expandafter\ifx\csname bibnamefont\endcsname\relax
  \def\bibnamefont#1{#1}\fi
\expandafter\ifx\csname bibfnamefont\endcsname\relax
  \def\bibfnamefont#1{#1}\fi
\expandafter\ifx\csname citenamefont\endcsname\relax
  \def\citenamefont#1{#1}\fi
\expandafter\ifx\csname url\endcsname\relax
  \def\url#1{\texttt{#1}}\fi
\expandafter\ifx\csname urlprefix\endcsname\relax\def\urlprefix{URL }\fi
\providecommand{\bibinfo}[2]{#2}
\providecommand{\eprint}[2][]{\url{#2}}

\bibitem[{\citenamefont{Goldstone}(1957)}]{Goldstone:1957zz}
\bibinfo{author}{\bibfnamefont{J.}~\bibnamefont{Goldstone}},
  \bibinfo{journal}{Proc.Roy.Soc.Lond.A Math.Phys.Eng.Sci.}
  \textbf{\bibinfo{volume}{239}}, \bibinfo{pages}{267} (\bibinfo{year}{1957}).

\bibitem[{\citenamefont{Moshinsky}(1958)}]{Moshinsky195819}
\bibinfo{author}{\bibfnamefont{M.}~\bibnamefont{Moshinsky}},
  \bibinfo{journal}{Nuclear Physics} \textbf{\bibinfo{volume}{8}},
  \bibinfo{pages}{19 } (\bibinfo{year}{1958}), ISSN \bibinfo{issn}{0029-5582}.

\bibitem[{\citenamefont{Skyrme}(1959)}]{Skyrme:1959zz}
\bibinfo{author}{\bibfnamefont{T.}~\bibnamefont{Skyrme}},
  \bibinfo{journal}{Nucl. Phys.} \textbf{\bibinfo{volume}{9}},
  \bibinfo{pages}{615} (\bibinfo{year}{1959}).

\bibitem[{\citenamefont{Vautherin and Brink}(1972)}]{Vautherin:1971aw}
\bibinfo{author}{\bibfnamefont{D.}~\bibnamefont{Vautherin}} \bibnamefont{and}
  \bibinfo{author}{\bibfnamefont{D.~M.} \bibnamefont{Brink}},
  \bibinfo{journal}{Phys. Rev.} \textbf{\bibinfo{volume}{C5}},
  \bibinfo{pages}{626} (\bibinfo{year}{1972}).

\bibitem[{\citenamefont{Bender et~al.}(2003)\citenamefont{Bender, Heenen, and
  Reinhard}}]{Bender:2003jk}
\bibinfo{author}{\bibfnamefont{M.}~\bibnamefont{Bender}},
  \bibinfo{author}{\bibfnamefont{P.-H.} \bibnamefont{Heenen}},
  \bibnamefont{and} \bibinfo{author}{\bibfnamefont{P.-G.}
  \bibnamefont{Reinhard}}, \bibinfo{journal}{Rev. Mod. PHys.}
  \textbf{\bibinfo{volume}{75}}, \bibinfo{pages}{121} (\bibinfo{year}{2003}).

\bibitem[{\citenamefont{Dutra et~al.}(2012)\citenamefont{Dutra, Lourenco,
  Sa~Martins, Delfino, Stone et~al.}}]{Dutra:2012mb}
\bibinfo{author}{\bibfnamefont{M.}~\bibnamefont{Dutra}},
  \bibinfo{author}{\bibfnamefont{O.}~\bibnamefont{Lourenco}},
  \bibinfo{author}{\bibfnamefont{J.}~\bibnamefont{Sa~Martins}},
  \bibinfo{author}{\bibfnamefont{A.}~\bibnamefont{Delfino}},
  \bibinfo{author}{\bibfnamefont{J.}~\bibnamefont{Stone}},
  \bibnamefont{et~al.}, \bibinfo{journal}{Phys.Rev.}
  \textbf{\bibinfo{volume}{C85}}, \bibinfo{pages}{035201}
  (\bibinfo{year}{2012}), \eprint{1202.3902}.

\bibitem[{\citenamefont{Ruiz~Arriola}(2010)}]{Arriola:2010hj}
\bibinfo{author}{\bibfnamefont{E.}~\bibnamefont{Ruiz~Arriola}}
  (\bibinfo{year}{2010}), \eprint{nucl-th/1009.4161}.

\bibitem[{\citenamefont{Nakamura}(2004)}]{Nakamura:2004ek}
\bibinfo{author}{\bibfnamefont{S.~X.}~\bibnamefont{Nakamura}}
  \bibinfo{journal}{Prog.Theor.Phys.} \textbf{\bibinfo{volume}{114}},
  \bibinfo{pages}{77} (\bibinfo{year}{2005}{\natexlab{a}}),



\bibitem[{\citenamefont{Navarro~Perez
  et~al.}(2012{\natexlab{a}})\citenamefont{Navarro~Perez, Amaro, and
  Ruiz~Arriola}}]{NavarroPerez:2011fm}
\bibinfo{author}{\bibfnamefont{R.}~\bibnamefont{Navarro~Perez}},
  \bibinfo{author}{\bibfnamefont{J.}~\bibnamefont{Amaro}}, \bibnamefont{and}
  \bibinfo{author}{\bibfnamefont{E.}~\bibnamefont{Ruiz~Arriola}},
  \bibinfo{journal}{Prog.Part.Nucl.Phys.} \textbf{\bibinfo{volume}{67}},
  \bibinfo{pages}{359} (\bibinfo{year}{2012}{\natexlab{a}}),
  \eprint{1111.4328}.

\bibitem[{\citenamefont{Navarro~Perez
  et~al.}(2012{\natexlab{b}})\citenamefont{Navarro~Perez, Amaro, and
  Ruiz~Arriola}}]{NavarroPerez:2012qr}
\bibinfo{author}{\bibfnamefont{R.}~\bibnamefont{Navarro~Perez}},
  \bibinfo{author}{\bibfnamefont{J.}~\bibnamefont{Amaro}}, \bibnamefont{and}
  \bibinfo{author}{\bibfnamefont{E.}~\bibnamefont{Ruiz~Arriola}},
  \bibinfo{journal}{Few Body Syst.} \textbf{\bibinfo{volume}{54}},
  \bibinfo{pages}{1487} (\bibinfo{year}{2013}).

\bibitem[{\citenamefont{Bogner et~al.}(2003{\natexlab{a}})\citenamefont{Bogner,
  Kuo, Schwenk, Entem, and Machleidt}}]{Bogner:2001gq}
\bibinfo{author}{\bibfnamefont{S.~K.} \bibnamefont{Bogner}},
  \bibinfo{author}{\bibfnamefont{T.~T.~S.} \bibnamefont{Kuo}},
  \bibinfo{author}{\bibfnamefont{A.}~\bibnamefont{Schwenk}},
  \bibinfo{author}{\bibfnamefont{D.~R.} \bibnamefont{Entem}}, \bibnamefont{and}
  \bibinfo{author}{\bibfnamefont{R.}~\bibnamefont{Machleidt}},
  \bibinfo{journal}{Phys. Lett.} \textbf{\bibinfo{volume}{B576}},
  \bibinfo{pages}{265} (\bibinfo{year}{2003}{\natexlab{a}}),

\bibitem[{\citenamefont{Bogner et~al.}(2003{\natexlab{b}})\citenamefont{Bogner,
  Kuo, and Schwenk}}]{Bogner:2003wn}
\bibinfo{author}{\bibfnamefont{S.~K.} \bibnamefont{Bogner}},
  \bibinfo{author}{\bibfnamefont{T.~T.~S.} \bibnamefont{Kuo}},
  \bibnamefont{and} \bibinfo{author}{\bibfnamefont{A.}~\bibnamefont{Schwenk}},
  \bibinfo{journal}{Phys. Rept.} \textbf{\bibinfo{volume}{386}},
  \bibinfo{pages}{1} (\bibinfo{year}{2003}{\natexlab{b}}).

\bibitem[{\citenamefont{Bogner et~al.}(2007{\natexlab{a}})\citenamefont{Bogner,
  Furnstahl, and Perry}}]{Bogner:2006pc}
\bibinfo{author}{\bibfnamefont{S.}~\bibnamefont{Bogner}},
  \bibinfo{author}{\bibfnamefont{R.}~\bibnamefont{Furnstahl}},
  \bibnamefont{and} \bibinfo{author}{\bibfnamefont{R.}~\bibnamefont{Perry}},
  \bibinfo{journal}{Phys.Rev.} \textbf{\bibinfo{volume}{C75}},
  \bibinfo{pages}{061001} (\bibinfo{year}{2007}{\natexlab{a}}).

\bibitem[{\citenamefont{Coraggio et~al.}(2009)\citenamefont{Coraggio, Covello,
  Gargano, Itaco, and Kuo}}]{Coraggio:2008in}
\bibinfo{author}{\bibfnamefont{L.}~\bibnamefont{Coraggio}},
  \bibinfo{author}{\bibfnamefont{A.}~\bibnamefont{Covello}},
  \bibinfo{author}{\bibfnamefont{A.}~\bibnamefont{Gargano}},
  \bibinfo{author}{\bibfnamefont{N.}~\bibnamefont{Itaco}}, \bibnamefont{and}
  \bibinfo{author}{\bibfnamefont{T.~T.~S.} \bibnamefont{Kuo}},
  \bibinfo{journal}{Prog. Part. Nucl. Phys.} \textbf{\bibinfo{volume}{62}},
  \bibinfo{pages}{135} (\bibinfo{year}{2009}), \eprint{0809.2144}.

\bibitem[{\citenamefont{Bogner et~al.}(2010)\citenamefont{Bogner, Furnstahl,
  and Schwenk}}]{Bogner:2009bt}
\bibinfo{author}{\bibfnamefont{S.~K.} \bibnamefont{Bogner}},
  \bibinfo{author}{\bibfnamefont{R.~J.} \bibnamefont{Furnstahl}},
  \bibnamefont{and} \bibinfo{author}{\bibfnamefont{A.}~\bibnamefont{Schwenk}},
  \bibinfo{journal}{Prog. Part. Nucl. Phys.} \textbf{\bibinfo{volume}{65}},
  \bibinfo{pages}{94} (\bibinfo{year}{2010}), \eprint{0912.3688}.

\bibitem[{\citenamefont{Furnstahl and Hebeler}(2013)}]{Furnstahl:2013oba}
\bibinfo{author}{\bibfnamefont{R.}~\bibnamefont{Furnstahl}} \bibnamefont{and}
  \bibinfo{author}{\bibfnamefont{K.}~\bibnamefont{Hebeler}}
  (\bibinfo{year}{2013}), \eprint{1305.3800}.

\bibitem[{\citenamefont{Holt et~al.}(2004)\citenamefont{Holt, Kuo, Brown, and
  Bogner}}]{Holt:2003rj}
\bibinfo{author}{\bibfnamefont{J.~D.} \bibnamefont{Holt}},
  \bibinfo{author}{\bibfnamefont{T.~T.~S.} \bibnamefont{Kuo}},
  \bibinfo{author}{\bibfnamefont{G.~E.} \bibnamefont{Brown}}, \bibnamefont{and}
  \bibinfo{author}{\bibfnamefont{S.~K.} \bibnamefont{Bogner}},
  \bibinfo{journal}{Nucl. Phys.} \textbf{\bibinfo{volume}{A733}},
  \bibinfo{pages}{153} (\bibinfo{year}{2004}), \eprint{nucl-th/0308036}.

\bibitem[{\citenamefont{Anderson et~al.}(2008)}]{Anderson:2008mu}
\bibinfo{author}{\bibfnamefont{E.}~\bibnamefont{Anderson}}
  \bibnamefont{et~al.}, \bibinfo{journal}{Phys. Rev.}
  \textbf{\bibinfo{volume}{C77}}, \bibinfo{pages}{037001}
  (\bibinfo{year}{2008}), \eprint{0801.1098}.

\bibitem[{\citenamefont{Szpigel and Perry}(1999)}]{Szpigel:1999gf}
\bibinfo{author}{\bibfnamefont{S.}~\bibnamefont{Szpigel}} \bibnamefont{and}
  \bibinfo{author}{\bibfnamefont{R.~J.} \bibnamefont{Perry}}
  (\bibinfo{year}{1999}), \eprint{nucl-th/9906031}.

\bibitem[{\citenamefont{Harada et~al.}(2006)\citenamefont{Harada, Inoue, and
  Kubo}}]{Harada:2005tw}
\bibinfo{author}{\bibfnamefont{K.}~\bibnamefont{Harada}},
  \bibinfo{author}{\bibfnamefont{K.}~\bibnamefont{Inoue}}, \bibnamefont{and}
  \bibinfo{author}{\bibfnamefont{H.}~\bibnamefont{Kubo}},
  \bibinfo{journal}{Phys.Lett.} \textbf{\bibinfo{volume}{B636}},
  \bibinfo{pages}{305} (\bibinfo{year}{2006}).

\bibitem[{\citenamefont{Entem et~al.}(2008)\citenamefont{Entem, Ruiz~Arriola,
  Pavon~Valderrama, and Machleidt}}]{Entem:2007jg}
\bibinfo{author}{\bibfnamefont{D.}~\bibnamefont{Entem}},
  \bibinfo{author}{\bibfnamefont{E.}~\bibnamefont{Ruiz~Arriola}},
  \bibinfo{author}{\bibfnamefont{M.}~\bibnamefont{Pavon~Valderrama}},
  \bibnamefont{and}
  \bibinfo{author}{\bibfnamefont{R.}~\bibnamefont{Machleidt}},
  \bibinfo{journal}{Phys.Rev.} \textbf{\bibinfo{volume}{C77}},
  \bibinfo{pages}{044006} (\bibinfo{year}{2008}).

\bibitem[{\citenamefont{Perez et~al.}(2013)\citenamefont{Perez, Amaro, and
  Arriola}}]{Perez:2013mwa}
\bibinfo{author}{\bibfnamefont{R.} \bibnamefont{Navarro~Perez}},
  \bibinfo{author}{\bibfnamefont{J.}~\bibnamefont{Amaro}}, \bibnamefont{and}
  \bibinfo{author}{\bibfnamefont{E.} \bibnamefont{Ruiz~Arriola}}
  \bibinfo{journal}{Phys.Rev.} \textbf{\bibinfo{volume}{C88}},
  \bibinfo{pages}{024002} (\bibinfo{year}{2013}).

\bibitem[{\citenamefont{Glazek and Wilson}(1993)}]{Glazek:1993rc}
\bibinfo{author}{\bibfnamefont{S.~D.} \bibnamefont{Glazek}} \bibnamefont{and}
  \bibinfo{author}{\bibfnamefont{K.~G.} \bibnamefont{Wilson}},
  \bibinfo{journal}{Phys. Rev.} \textbf{\bibinfo{volume}{D48}},
  \bibinfo{pages}{5863} (\bibinfo{year}{1993}).

\bibitem[{\citenamefont{Glazek and Wilson}(1994)}]{Glazek:1994qc}
\bibinfo{author}{\bibfnamefont{S.~D.} \bibnamefont{Glazek}} \bibnamefont{and}
  \bibinfo{author}{\bibfnamefont{K.~G.} \bibnamefont{Wilson}},
  \bibinfo{journal}{Phys. Rev.} \textbf{\bibinfo{volume}{D49}},
  \bibinfo{pages}{4214} (\bibinfo{year}{1994}).

\bibitem[{\citenamefont{Wegner}(2001)}]{Wegner200177}
\bibinfo{author}{\bibfnamefont{F.~J.} \bibnamefont{Wegner}},
  \bibinfo{journal}{Physics Reports} \textbf{\bibinfo{volume}{348}},
  \bibinfo{pages}{77 } (\bibinfo{year}{2001}).

\bibitem[{\citenamefont{Kehrein}(2006)}]{Kehrein:2006ti}
\bibinfo{author}{\bibfnamefont{S.}~\bibnamefont{Kehrein}},
  \emph{\bibinfo{title}{{The flow equation approach to many-particle systems}}}
  (\bibinfo{publisher}{Springer}, \bibinfo{year}{2006}).

\bibitem[{\citenamefont{Calle~Cordon and
  Ruiz~Arriola}(2008)}]{CalleCordon:2008cz}
\bibinfo{author}{\bibfnamefont{A.}~\bibnamefont{Calle~Cordon}}
  \bibnamefont{and}
  \bibinfo{author}{\bibfnamefont{E.}~\bibnamefont{Ruiz~Arriola}},
  \bibinfo{journal}{Phys. Rev.} \textbf{\bibinfo{volume}{C78}},
  \bibinfo{pages}{054002} (\bibinfo{year}{2008}).

\bibitem[{\citenamefont{Timoteo et~al.}(2012)\citenamefont{Timoteo, Szpigel,
  and Ruiz~Arriola}}]{Timoteo:2011tt}
\bibinfo{author}{\bibfnamefont{V.S.}~\bibnamefont{Tim\'oteo}},
  \bibinfo{author}{\bibfnamefont{S.}~\bibnamefont{Szpigel}}, \bibnamefont{and}
  \bibinfo{author}{\bibfnamefont{E.}~\bibnamefont{Ruiz~Arriola}},
  \bibinfo{journal}{Phys.Rev.} \textbf{\bibinfo{volume}{C86}},
  \bibinfo{pages}{034002} (\bibinfo{year}{2012}), \eprint{1108.1162}.

\bibitem[{\citenamefont{Arriola et~al.}(2013)\citenamefont{Arriola, Timoteo,
  and Szpigel}}]{Arriola:2013nja}
\bibinfo{author}{\bibfnamefont{E.}~\bibnamefont{Ruiz~Arriola}},
  \bibinfo{author}{\bibfnamefont{V.~S.}~\bibnamefont{Tim\'oteo}}, \bibnamefont{and}
  \bibinfo{author}{\bibfnamefont{S.}~\bibnamefont{Szpigel}}, 
  \bibinfo{journal}{PoS CD} \textbf{\bibinfo{volume}{12}},
  \bibinfo{pages}{106}
  (\bibinfo{year}{2013}), \eprint{nucl-th/1302.3978}.


\bibitem[{\citenamefont{Amghar and Desplanques}(1995)}]{Amghar:1995av}
\bibinfo{author}{\bibfnamefont{A.}~\bibnamefont{Amghar}} \bibnamefont{and}
  \bibinfo{author}{\bibfnamefont{B.}~\bibnamefont{Desplanques}},
  \bibinfo{journal}{Nucl.Phys.} \textbf{\bibinfo{volume}{A585}},
  \bibinfo{pages}{657} (\bibinfo{year}{1995}).

\bibitem[{\citenamefont{Szpigel et~al.}(2011)\citenamefont{Szpigel, Timoteo,
  and Duraes}}]{Szpigel:2010bj}
\bibinfo{author}{\bibfnamefont{S.}~\bibnamefont{Szpigel}},
  \bibinfo{author}{\bibfnamefont{V.~S.} \bibnamefont{Tim\'oteo}},
  \bibnamefont{and} \bibinfo{author}{\bibfnamefont{F.~d.~O.}
  \bibnamefont{Duraes}}, \bibinfo{journal}{Annals Phys.}
  \textbf{\bibinfo{volume}{326}}, \bibinfo{pages}{364} (\bibinfo{year}{2011}),
  \eprint{1003.4663}.

\bibitem[{\citenamefont{Bogner et~al.}(2007{\natexlab{b}})\citenamefont{Bogner,
  Furnstahl, Ramanan, and Schwenk}}]{Bogner:2006vp}
\bibinfo{author}{\bibfnamefont{S.}~\bibnamefont{Bogner}},
  \bibinfo{author}{\bibfnamefont{R.}~\bibnamefont{Furnstahl}},
  \bibinfo{author}{\bibfnamefont{S.}~\bibnamefont{Ramanan}}, \bibnamefont{and}
  \bibinfo{author}{\bibfnamefont{A.}~\bibnamefont{Schwenk}},
  \bibinfo{journal}{Nucl.Phys.} \textbf{\bibinfo{volume}{A784}},
  \bibinfo{pages}{79} (\bibinfo{year}{2007}{\natexlab{b}}),
  \eprint{nucl-th/0609003}.

\bibitem[{\citenamefont{Harada et~al.}(2011)\citenamefont{Harada, Kubo, and
  Yamamoto}}]{Harada:2010ba}
\bibinfo{author}{\bibfnamefont{K.}~\bibnamefont{Harada}},
  \bibinfo{author}{\bibfnamefont{H.}~\bibnamefont{Kubo}}, \bibnamefont{and}
  \bibinfo{author}{\bibfnamefont{Y.}~\bibnamefont{Yamamoto}},
  \bibinfo{journal}{Phys.Rev.} \textbf{\bibinfo{volume}{C83}},
  \bibinfo{pages}{034002} (\bibinfo{year}{2011}).

\end{thebibliography}

\end{document}